\documentclass[a4paper,11pt]{article}
\usepackage{array}
\usepackage{appendix} % 加载宏包
\usepackage{jheppub} % for details on the use of the package, please
\usepackage{setspace,amsmath,bm,amssymb,graphicx,mathrsfs,footmisc,multirow,booktabs}
\usepackage{braket}
\usepackage{enumitem}
\usepackage{datetime, scrtime}
\usepackage{subfigure}
\usepackage{float}
\usepackage{soul}
\usepackage{tikz}
\usepackage{xcolor, color, framed}
\usepackage{url}
\usepackage{array}

\hypersetup{
    colorlinks=true,
    linkcolor=red,
    anchorcolor=green,
    citecolor=blue
}

\usepackage[normalem]{ulem} 
\usepackage[font={footnotesize}]{caption} 
\usepackage{placeins} 
\usepackage{dcolumn}
\usepackage{lineno}
\usepackage{siunitx}
\graphicspath{{./figures/}}

\title{Unified Extraction of In-Medium Heavy Quark Potentials from RHIC to LHC Energies via Deep Learning}

 \author[a]{Jiamin Liu}\emailAdd{liujiamin@tju.edu.cn}
  \affiliation[a]{Department of Physics, Tianjin University, Tianjin 300354, China }
  
 \author[b, c]{Kai Zhou}\emailAdd{zhoukai@cuhk.edu.cn}
   \affiliation[b]{School of Science and Engineering, The Chinese University of Hong Kong
(Shenzhen), Longgang, Shenzhen, Guangdong, 518172, China}
   \affiliation[c]{School of Artificial Intelligence, The Chinese University of Hong Kong
(Shenzhen), Longgang, Shenzhen, Guangdong, 518172, China}

 \author[a]{Baoyi Chen}\emailAdd{baoyi.chen@tju.edu.cn}
\date{\today}

\abstract{
 We use deep learning under Bayesian perspective 
 to quantitatively extract the in-medium heavy quark (HQ) potential from bottomonium nuclear modification factors ($R_{AA}$) measured across multiple heavy ion collision systems at the Large Hadron Collider (LHC) and the Relativistic Heavy-Ion Collider (RHIC). The in-medium HQ potential, comprising both a real and imaginary part, is parameterized and incorporated into a time-dependent Schr\"odinger equation to model the wave function evolution of $b\bar{b}$ dipoles within a hydrodynamically evolving hot QCD medium. We construct Convolutional Neural Networks (CNNs) to capture the non-linear correspondence between the heavy quark potential $V(T,r)$ and the bottomonium $R_{AA}$ for Pb-Pb collisions at 5.02 TeV and 2.76 TeV, and Au-Au collisions at 200 GeV. Training datasets are generated by sampling the potential parameters and are further augmented using Principal Component Analysis (PCA) and Gaussian Process Regression (GPR). After validating the stability and correctness of the CNNs, we employ Stochastic Gradient Langevin Dynamics (SGLD) to perform a simultaneous Bayesian inverse extraction of the optimal potential parameters and their posterior distributions using experimental data of bottomonium $R_{AA}$ in both LHC and RHIC energies. Our joint multi-energy extraction suggests that, within the present parametrization and hydrodynamic background, the real part of the in-medium potential remains close to the vacuum Cornell form, corresponding to a relatively weak screened Debye mass across RHIC to LHC energies. By contrast, the imaginary part is more strongly constrained by the data and provides the dominant contribution to bottomonium suppression from RHIC to LHC energies.
}

\begin{document}
\maketitle
\flushbottom

\section{Introduction}

The theory of strong interactions predicts a deconfined state of matter composed of quarks and gluons, known as the quark-gluon plasma (QGP). This state can be created via a smooth crossover phase transition in high-energy nuclear collisions~\cite{Bazavov:2011nk}. Furthermore, Experimental measurements of light-hadron spectra suggest that the QGP behaves as a nearly perfect fluid with a minimal ratio of the viscosity over entropy density ~\cite{ALICE:2010suc,STAR:2017sal}, a behavior that is effectively captured by hydrodynamic frameworks~\cite{Pang:2012he,Gale:2013da,Shen:2014vra}. Heavy quarks, primarily produced during initial hard parton scatterings, serve as ideal probes of the early stage of heavy-ion collisions. Notably, the anomalous suppression of heavy quarkonium was proposed nearly forty years ago as a definitive signature of the deconfined medium~\cite{Matsui:1986dk}.
Quarkonium yield suppression observed in relativistic heavy-ion collisions is predominantly attributed to two mechanisms: the color screening effect and inelastic scattering with thermal partons~\cite{Song:2025zfy,Chen:2018dqg,Du:2017qkv,Strickland:2011aa,Zhao:2011cv,Blaizot:2018oev,Yan:2006ve,Karsch:1987pv}. Numerous phenomenological frameworks, including transport models~\cite{Grandchamp:2003uw,Zhou:2014kka}, complex potential models~\cite{Krouppa:2017jlg,Wen:2022utn}, statistical hadronization models~\cite{Andronic:2006ky}, and semi-classical or open quantum system approaches~\cite{Yao:2020eqy,Brambilla:2020qwo,Delorme:2024rdo}, have been developed to study the nuclear modification factors and anisotropic momentum distributions of charmonium and bottomonium across energy scales ranging from the Relativistic Heavy Ion Collider (RHIC) to the Large Hadron Collider (LHC). Within these frameworks, the real part of the heavy quark potential is reduced by the color screening. The imaginary part of the potential comes from inelastic scatterings with thermal partons of QGP~\cite{Margotta:2011ta}.
Lattice QCD calculations yield distinct heavy quark potentials at finite temperatures, which vary substantially depending on the extraction methodology~\cite{Bala:2021fkm}. Consequently, it is imperative to quantitatively constrain these heavy quark potentials in the hot deconfined medium by leveraging the extensive experimental datasets available for quarkonium states.

In previous studies, Bayesian inference has been employed to quantitatively extract diffusion coefficients~\cite{Xu:2017obm} and in-medium heavy quark potentials~\cite{Zheng:2025bdc} from experimental data on open and hidden heavy-flavor hadrons, respectively. Complementing these methods, deep neural networks (DNNs) offer distinct advantages in capturing features within high-dimensional data in nuclear physics. For instance, various DNN architectures have been developed to clarify the order of phase transitions in heavy-ion collisions~\cite{Pang:2016vdc, OmanaKuttan:2020btb, Zhou:2023pti}, investigate jet quenching~\cite{Du:2021pqa,Yang:2022yfr}, explore nuclear structure and nucleon-nucleon correlations~\cite{Huang:2025uvc}, simulate the hydrodynamic evolution of the quark-gluon plasma~\cite{Huang:2018fzn}, reconstruct spectral functions~\cite{Wang:2021jou}, and quantify the diffusion coefficients of heavy quarks~\cite{Guo:2023phd}. In this work, we construct a deep neural network to map the correspondence between in-medium heavy quark potentials and bottomonium observables across various collision energies. Specifically, the parameterized heavy quark potentials serve as the input to the network, while the output is a multi-dimensional vector comprising bottomonium nuclear modification factors ($R_{AA}$) across various centralities, transverse momentum ($p_T$) bins, and collision energies ($\sqrt{s_{NN}}$). To incorporate more stringent constraints on the potential, experimental data from Pb-Pb collisions at 5.02 TeV and 2.76 TeV, as well as Au-Au collisions at 200 GeV, are integrated simultaneously. 

In the generation of the training datasets, the in-medium heavy quark potentials, comprising both real and imaginary components, are characterized using a multi-parameter representation. These potentials are then incorporated into a non-linear, time-dependent Schr\"odinger equation, which is coupled with the hydrodynamic model to determine the final nuclear modification factors ($R_{AA}$) for the $\Upsilon(1S)$, $\Upsilon(2S)$, and $\Upsilon(3S)$ states across various collision systems. By systematically varying the potential parameters, we generate a comprehensive dataset of diverse potentials and their corresponding bottomonium $R_{AA}$ values, which serves as the training set for our Convolutional Neural Networks (CNNs). To address the challenge of a limited number of samples, we employ Principal Component Analysis (PCA) and Gaussian Process Regression (GPR) for data augmentation, thereby enhancing the model’s generalization capability and its coverage of the parameter space. Once the CNNs are well-trained independently for collision systems (200 GeV, 2.76 TeV, 5.02 TeV), we incorporate experimental data points and utilize Stochastic Gradient Langevin Dynamics (SGLD) to perform a Bayesian-like inverse extraction. This process aims to determine the temperature and distance dependence of the heavy quark potential $V(r,T)$ that simultaneously describes all experimental observables. Finally, the posterior distributions of the potential parameters are obtained through the SGLD sampling process.

The remainder of this paper is organized as follows. Section II introduces the Schr\"odinger-based framework for bottomonium evolution, coupled with the hydrodynamic model; the parameterization of the in-medium heavy quark potential is also detailed.
Section III introduces the Convolutional neural network(CNN) surrogate model, the Principal Component Analysis (PCA)/Gaussian Process (GP) data augmentation strategy, and the SGLD-based Bayesian inference procedure. Section IV presents the extracted values and posterior distributions of the potential parameters inferred from experimental bottomonium data, alongside an analysis of the stability and validation of the CNN models. Finally,  Section V provides a summary of our findings.

\section{Complex potential model for bottomonium}

\subsection{Schr\"odinger equation}

Given the large mass of the bottom quark, relativistic effects in the internal evolution of the \( b\bar{b} \) dipole wave function can be safely neglected. The inner evolution of the bottomonium wave function in the hot QCD medium can be effectively studied using the time-dependent Schr\"odinger equation by incorporating appropriate in-medium HQ  potentials into the Hamiltonian of the \( b\bar{b} \) dipole~\cite{Brambilla:2020qwo,Zheng:2025bdc}. By considering the color singlet component of the wave function in the Schr\"odinger equation, the radial part of the \( b\bar{b} \) dipole wave function can be separated as follows, attributed to the spherically symmetric potential,
\begin{align}
i\hbar \frac{\partial}{\partial t} \psi(r, t) = 
\left( -\frac{\hbar^2}{2m_\mu} \frac{\partial^2}{\partial r^2} + V(r, T) + \frac{L(L + 1)\hbar^2}{2m_\mu r^2} \right) \psi(r, t),
\label{eq:3}
\end{align}
where $r$ is the distance between the bottom and anti-bottom quarks. \( t \) represents the proper time in the center-of-mass frame of the $b\bar b$ dipole. The reduced mass of the bottom quarks in the center-of-mass frame is defined as \( m_\mu = {m_1 m_2}/(m_1 + m_2) \), with the bottom quark mass \( m_b = 4.62 \, \text{GeV} \). The quantum number \( L = 0, 1, \dots \) corresponds to the angular momentum. The function \( \psi(r,t) = rR(r,t) \) is the product of the radius \( r \) and the radial part of the \( b\bar{b} \) dipole wave function. 
The wave packet \( \psi(r,t) \) can be expanded in terms of a complete set of vectors consisting of multiple bottomonium eigenstates 
$\frac{\psi(r,t)}{r} = \sum_{nl} c_{nl}(t) \Phi_{nl}(r)$,
where \( n \) and \( l \) represent the radial and angular quantum numbers, respectively. \( \Phi_{nl}(r) \) is defined as the eigenstate of the vacuum Cornell potential \( V_c(r) = -\frac{\alpha}{r} + \sigma r \), which has been parametrized to fit the masses of quarkonium vacuum eigenstates, with the values to be \( \alpha = \pi/12 \) and \( \sigma = 0.2 \, \text{(GeV/c)}^2 \)~\cite{Satz:2005hx}. The hot QCD medium effects are encoded in the heavy quark potential \( V(r,T) \), which changes the wave function of $b\bar b$ dipoles and the corresponding fractions of each bottomonium eigenstate, \( |c_{nl}(t)|^2 \). The values of \( |c_{nl}(t=0)|^2 \) and \( |c_{nl}(t=t_f)|^2 \) characterize the initial and final fractions of the bottomonium eigenstate labeled by \( (n,l) \) within one \( b\bar{b} \) dipole wave package. These values will be used to calculate the bottomonium nuclear modification factor \( R_{AA} \).

The radial Schr\"odinger equation in Eq.~(\ref{eq:3}) can be numerically solved with the Crank-Nicolson method~\cite{crank1947practical} or the tridiagonal matrix algorithm (TMA)\cite{Barata:2025htx}. In natural units with $\hbar = c = 1$, Eq.~(\ref{eq:3}) is discretized in both the temporal and spatial domains, yielding the following form,
\begin{equation}
\mathcal{ T}_j \psi_j^{n+1} - \psi_{j-1}^{n+1} - \psi_{j+1}^{n+1} = \Gamma_j, \quad 1 \leq j \leq N-1, 
\label{eq:6}
\end{equation}
which is rewritten in the form of the tridiagonal matrix: 
\begin{equation}
\begin{bmatrix}
\mathcal{ T}_1 & -1 & 0 &  \cdots & 0 \\
-1 & \mathcal{T}_2 & -1 &  \cdots & 0 \\
0 & -1 & \mathcal{ T}_3 &  \cdots & 0 \\
\vdots & \vdots & \vdots & \ddots & \vdots \\
0 & \cdots & 0 & -1 & \mathcal{T}_N
\end{bmatrix}
\label{eq:7}
\begin{bmatrix}
\psi_1^{n+1} \\
\psi_2^{n+1} \\
\psi_3^{n+1} \\
\vdots \\
\psi_N^{n+1}
\end{bmatrix}
=
\begin{bmatrix}
\Gamma_1^n \\
\Gamma_2^n \\
\Gamma_3^n \\
\vdots \\
\Gamma_N^n
\end{bmatrix},
\end{equation}
with the diagonal elements \( \mathcal{T}_j \) and the term \( \Gamma_j \) defined as,
\begin{align}
\label{eq:8}
\mathcal{T}_j &= \frac{2m_\mu (\Delta r)^2}{i \Delta t} + 2 + m_\mu (\Delta r)^2 \Lambda_j^{n+1},\\
\label{eq:9}
\Gamma_j &= \left( \frac{2m_\mu (\Delta r)^2}{i \Delta t} - 2 - m_\mu (\Delta r)^2 \Lambda_j^{n} \right) \psi_j^n + \psi_{j-1}^n + \psi_{j+1}^n.
\end{align}
Here, $i$ denotes the imaginary unit, while $j$ and $n$ serve as the indices for the discrete radius $r_j$ and time $t_n$, respectively. The wave function $\psi_j^n \equiv \psi(r_j, t_n)$ is defined on a grid with $r_j = r_0 + j \Delta r$ and $t_n = t_0 + n \Delta t$. For the numerical simulations, the step sizes are chosen as $\Delta r = 0.03$~fm and $\Delta t = 0.02$~fm/$c$. To circumvent the numerical singularity associated with the $1/r$ term in the potential, a small offset $r_0 = 10^{-5}$~fm is introduced. The initial time is set to $t_0 = 0.6$~fm/$c$, consistent with the typical initialization time of the bulk medium evolution in hydrodynamic frameworks~\cite{Shen:2010uy}. Detailed parameterizations of the in-medium HQ potential $V(r, T)$ will be discussed in the subsequent section.

The numerical solution of the Schr\"odinger equation also requires specified initial conditions, $\psi(r, t_0)$. In relativistic nuclear collisions, $b\bar{b}$ dipoles are produced during initial hard parton scatterings on an extremely short time scale, $\tau_{b\bar{b}} \sim 1/(2m_b) \sim 0.02$~fm/$c$. These $b\bar{b}$ dipoles subsequently evolve from a compact wave packet toward specific bottomonium eigenstates, a mechanism known as the formation process~\cite{Song:2015bja}. Based on the uncertainty principle, the formation time scale varies across bottomonium eigenstates, and is estimated to be comparable to the thermalization time of the bulk medium, $t_0$. For simplicity, we initialize the $b\bar{b}$ wave function at $t = t_0$ as a bottomonium eigenstate, such as $\Upsilon(1S)$, $\Upsilon(2S)$, $\chi_b(1P)$, $\Upsilon(3S)$, or $\chi_b(2P)$~\cite{Wen:2022yjx}. For $t \geq t_0$, the subsequent evolution of the wave function is governed by the time-dependent Schr\"odinger equation within the expanding medium.

In relativistic nuclear collisions, multiple $b\bar{b}$ dipoles are produced throughout the transverse overlap area, each carrying a distinct total momentum. These $b\bar{b}$ dipoles propagate through the QGP along straight-line trajectories at constant velocities, with their positions updated as ${\bf x}_\Upsilon(t) = {\bf x}_\Upsilon(t_0) + \frac{{\bf p}_\Upsilon}{E_\Upsilon} (t - t_0)$, where ${\bf p}_\Upsilon$ and $E_\Upsilon = \sqrt{m_\Upsilon^2 + |{\bf p}_\Upsilon|^2}$ denote the momentum and total energy of the $b\bar{b}$ dipole, respectively. Notably, the energy loss of color-singlet bottomonium states within the QGP is neglected, in contrast to that of single heavy quarks. The initial positions and velocities of the $b\bar{b}$ dipoles determine the local medium temperatures experienced along their paths before they exit the deconfined medium. The initial transverse positions are sampled according to the binary collision density, $dN_\Upsilon/d{\bf x}_T \propto n_{coll}({\bf x}_T)$. Once a $b\bar{b}$ dipole leaves the hot medium, the evolution governed by the in-medium Schr\"odinger equation ceases, and the heavy quark potential reverts to the vacuum Cornell potential.

To account for the diverse production points and kinematic distributions, we sample a large ensemble of $b\bar{b}$ dipoles with varying initial positions and momenta, evolving each dipole on an event-by-event basis via the Schr\"odinger equation. By performing an ensemble average, we obtain the mean fractions $\langle |c_{nl}(t)|^2 \rangle_{\text{en}}$ for each bottomonium eigenstate within a single $b\bar{b}$ dipole. These quantities characterize the integrated effects of the hot medium on bottomonium survival and transition probabilities,
\begin{align}
\langle |c_{nl}(t)|^2 \rangle_{\text{en}} &=
\frac{
\int d{\bf x}_{\Upsilon}\, d{\bf p}_{\Upsilon}\,
|c_{nl}(t,{\bf x}_{\Upsilon},{\bf p}_{\Upsilon})|^{2}
\frac{dN_{AA}^{\Upsilon}}{d{\bf x}_{\Upsilon}\,d{\bf p}_{\Upsilon}}
}{
\int d{\bf x}_{\Upsilon} d{\bf p}_{\Upsilon}
\frac{dN_{AA}^{\Upsilon}}{d{\bf x}_{\Upsilon} d{\bf p}_{\Upsilon}}
}.
\label{eq:11}
\end{align}
Here, ${\bf x}_\Upsilon$ and ${\bf p}_\Upsilon$ denote the initial position and momentum of each $b\bar{b}$ dipole, which are sampled according to the factorized distribution $\frac{dN_\Upsilon}{d{\bf x}_\Upsilon d{\bf p}_\Upsilon} = \frac{dN_\Upsilon}{d{\bf x}_\Upsilon} \cdot \frac{dN_\Upsilon}{d{\bf p}_\Upsilon}$. The specific functional forms for the spatial distribution $\frac{dN_\Upsilon}{d{\bf x}_\Upsilon}$ and the momentum distribution $\frac{dN_\Upsilon}{d{\bf p}_\Upsilon}$ are detailed in the following sections.

\subsection{Initial distribution of bottomonium}

The final prompt production of $\Upsilon(1S)$ observed in experiments comprises both direct $\Upsilon(1S)$ production and feed-down contributions from higher-mass excited states. The initial yield ratios of these states in nuclear collisions are determined by their direct production cross sections $\sigma_{\text{direct}}(1S): \sigma_{\text{direct}}(1P): \sigma_{\text{direct}}(2S): \sigma_{\text{direct}}(2P): \sigma_{\text{direct}}(3S)$. Given that experimental prompt production cross sections, $\sigma_{\text{exp}}$, for various bottomonium states in $pp$ collisions have been measured~\cite{CMS:2016rpc,collaboration2019measurement, collaboration2014study, collaboration2015measurement, lansberg2020new} (see Table~\ref{pp-cross-section}), the direct cross sections can be extracted by incorporating the known branching ratios for the decay of excited states into lower-lying states~\cite{Brambilla:2020qwo,Zheng:2025bdc}. In Table~\ref{pp-cross-section}, the $\chi_b(1P)$ state represents the aggregate of the $\chi_{b0}(1P)$, $\chi_{b1}(1P)$, and $\chi_{b2}(1P)$ triplet. Similarly, $\chi_b(2P)$ denotes the sum of the $\chi_{b0}(2P)$, $\chi_{b1}(2P)$, and $\chi_{b2}(2P)$ states. In our Schr\"odinger model, we do not distinguish between the individual $J$ states of the $\chi_b$ triplets; instead, we treat them as combined $\chi_b(nP)$ states within the coupled evolution of the $\Upsilon(1S, 1P, 2S, 2P, 3S)$ system.
\begin{table}[h!]
\centering
\caption{Bottomonium production cross sections in the central rapidity bins in $\sqrt{s_{NN}} = 5.02$~TeV pp collision.}
\label{pp-cross-section}
\begin{tabular}{|c|c|c|c|c|c|}
\hline
\textbf{State} & $\Upsilon(1\mathrm{S})$ & 
$\chi_b(1\mathrm{P})$ & $\Upsilon(2\mathrm{S})$ & $\chi_b(2\mathrm{P})$ &
$\Upsilon(3\mathrm{S})$ \\
\hline
$\sigma_{\text{exp}}$ (nb) & 57.6 & 33.51 & 19 & 29.42 & 6.8 \\
$\sigma_{\text{direct}}$ (nb) & 37.97 & 44.2 & 18.27 & 37.68  & 8.21 \\
\hline
\end{tabular}
\end{table}

For Pb-Pb collisions at $\sqrt{s_{NN}} = 2.76$~TeV, the measured prompt production cross sections for $\Upsilon(1S)$ and $\Upsilon(2S)$ in the central rapidity region are $d\sigma/dy = 30.3$~nb and $10.0$~nb, respectively~\cite{CMS:2016rpc,Du:2017qkv}. Under the assumption that the production ratios among different bottomonium states remain invariant across varying collision energies, the prompt cross sections for the $\Upsilon(1P, 2P, 3S)$ states at 2.76~TeV can be inferred using the corresponding yield ratios measured at 5.02~TeV. Similarly, for Au-Au collisions at $\sqrt{s_{NN}} = 200$~GeV, according to the measurements from PHENIX~\cite{PHENIX:2012xws} and STAR~\cite{STAR:2013kwk} Collaboration, we extract the production cross sections for $\Upsilon(1S)$ and $\Upsilon(2S)$ in the central rapidity region as $d\sigma/dy = 2.35$~nb and $0.77$~nb, respectively. Following the same methodology, leveraging the yield ratios of bottomonium eigenstates measured in 5.02~TeV $pp$ collisions, we estimate the cross section for the $\Upsilon(3S)$ state at 200~GeV to be $d\sigma/dy = 0.27$~nb.

Regarding the initial spatial distribution of bottomonium, it is well known that the shadowing effect can modify the gluon density within the nucleus, thereby significantly influencing bottomonium production. To account for these cold nuclear matter effects, we incorporate spatial corrections using the shadowing factors calculated via the EPS09 NLO package~\cite{eskola2009eps09}. Consequently, the initial spatial distribution of bottomonium in nuclear collisions, which serves as the probability density for sampling the initial positions of $b\bar{b}$ dipoles, is expressed as:
\begin{align}
    dN_\Upsilon/d{\bf x}_T \propto \mathcal{R}_S({\bf x}_T, {\bf p}_T) T_A({\bf x}_T + {\bf b}/2) T_B({\bf x}_T - {\bf b}/2)
\end{align}
Here, $\mathcal{R}_S$ denotes the shadowing factor obtained from the EPS09 NLO model, which depends on both the spatial position within the nucleus and the transverse momentum of the produced bottomonium~\cite{Zhou:2014kka}. The nuclear thickness function, defined as $T_{A(B)}({\bf x}_T) = \int dz \rho({\bf x}_T, z)$~\cite{miller2007glauber}, is calculated by integrating the nuclear density over the longitudinal coordinate $z$. This density $\rho$ is assumed to follow the Woods-Saxon distribution. Additionally, ${\bf b}$ represents the impact parameter of the collision.

\begin{figure}[htbp]
\centering
\includegraphics[width=0.6\linewidth]{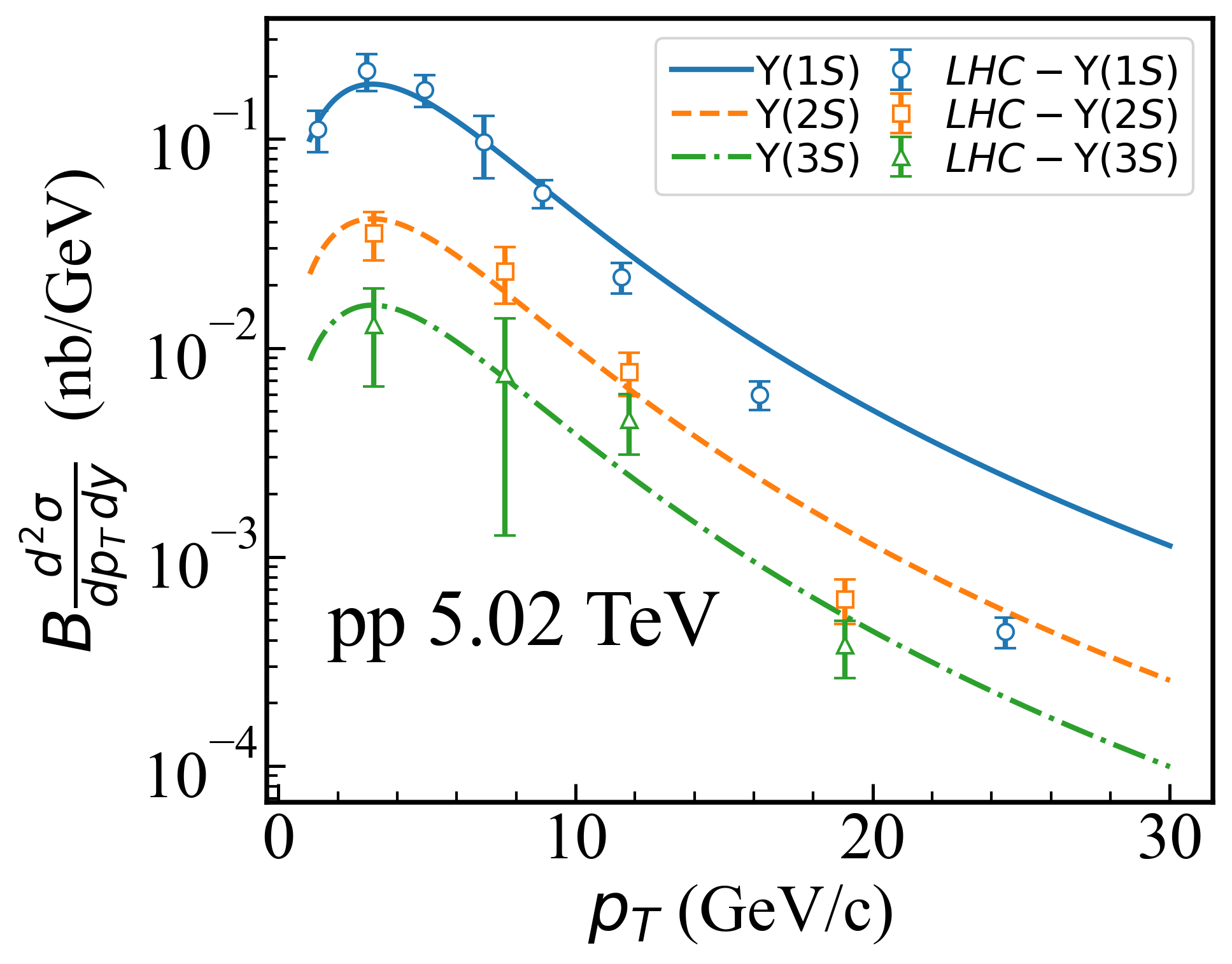}
  \caption{{Transverse-momentum ($p_T$) spectra of $\Upsilon(1S)$, $\Upsilon(2S)$, and $\Upsilon(3S)$ in $pp$ collisions at $\sqrt{s}=5.02~\mathrm{TeV}$ for the rapidity interval $0.00<|y|<0.75$~\cite{ATLAS:2017prf}. Data points represent $\mathcal{B}\, d^{2}\sigma/(dp_T\,dy)$ (nb/GeV), where $\mathcal{B}$ is the dimuon branching fraction. The solid curves show $Q_{\mathrm{th}}(p_T)=\mathcal{B}\,(d\sigma/dy)\,2\pi p_T\,F(p_T)$, where $F(p_T)$ is the unit-normalized shape from Eq.~(2.8), and $d\sigma/dy$ sets the overall normalization.}
 }
  \label{fig:pp-pT}
\end{figure}

The normalized initial transverse momentum distribution of the bottomonium ground state $\Upsilon(1S)$ in pp collisions can be fitted with the following form, 
\begin{align}
\frac{d N_{\Upsilon}}{2 \pi p_T dp_T} = 
\frac{(n - 1)}{\pi (n - 2) \langle p_T^2 \rangle_{pp}} 
\left[ 1 + \frac{p_T^2}{(n - 2) \langle p_T^2 \rangle_{pp}} \right]^{-n} ,
\label{eq:pp-pT}
\end{align}
where $n=2.5$ and $\langle p_T^2\rangle_{pp}$ represents the mean transverse momentum square of the bottomonium ground state in $pp$ collisions. The values are determined to be $\langle p_T^2 \rangle_{pp} = (80, 55, 28)~\text{GeV}^2/c^2$~\cite{Brambilla:2020qwo, thelhcbcollaboration2012measurement, Wen:2022yjx} by fitting the $\Upsilon(1S)$ spectra measured in central-rapidity $pp$ collisions at $\sqrt{s_{NN}} = 5.02$~TeV, $2.76$~TeV, and $200$~GeV, respectively. Given that the mass differences between various bottomonium eigenstates are small relative to the mean transverse momentum, the normalized initial $p_T$ distributions for both the excited states and the $b\bar{b}$ dipoles are assumed to be identical to those of the ground state. Their transverse momentum distributions in $5.02$~TeV $pp$ collisions are illustrated in Figure ~\ref{fig:pp-pT}. With the spatial and momentum distributions given above, one can sample the initial positions and momenta of $b\bar b$ dipoles event-by-event. After the $b\bar{b}$ dipoles exit the hot medium, the integrated effects of the in-medium heavy quark potentials are encoded in the final survival fractions of the bottomonium eigenstates within the dipoles. Finally, accounting for feed-down contributions from higher excited states, the prompt nuclear modification factor for $\Upsilon(1S)$ is given by~\cite{Wen:2022yjx},
\begin{align}
R_{AA}\bigl(\Upsilon(1\mathrm{S})\bigr) &=
\frac{
\sum_{n,l}
\langle |c_{nl}(t)|^{2}\rangle_{\text{en}}
f^{nl}_{\rm init} B_{nl\to1\mathrm{S}}
}{
\sum_{n,l}
\langle |c_{nl}(t_0)|^{2}\rangle_{\text{en}}
f^{nl}_{\rm init} B_{nl\to1\mathrm{S}}
},
\label{eq:12}
\end{align}
where $B_{nl \to 1S}$ denotes the branching ratio from the excited eigenstate labeled by $(n, l)$ to the ground state $\Upsilon(1S)$~\cite{Brambilla:2020qwo,Zheng:2025bdc}. The parameter $f_{\text{init}}^{nl}$ represents the initial direct production yield of the bottomonium eigenstate $(n, l)$ with a certain rapidity and transverse momentum. For other bottomonium states, such as $\Upsilon(2S)$ and $\Upsilon(3S)$, the prompt nuclear modification factors can be evaluated using a similar procedure.

\subsection{Parametrization of in-medium heavy quark potential}

The anomalous suppression of heavy quarkonium production is intrinsically linked to the in-medium potentials of heavy quarks, which govern the dynamical evolution of the quarkonium wave function within the deconfined medium. The medium-induced effects are encoded in the effective Hamiltonian of the heavy quark dipole. In this work, we focus exclusively on color-singlet states; accordingly, the dipole wave function in the Schr\"odinger equation is represented as a superposition of color-singlet eigenstates, while color-octet components are omitted. Consequently, quarkonium dissociation is effectively captured by the imaginary component of the potential, which leads to a reduction in the normalization of the $b\bar{b}$ wave function. Furthermore, the transition from octet to singlet states is neglected in this framework, as its contribution is expected to be minimal~\cite{Akamatsu:2014qsa}.

The dissociation rate of quarkonium, induced by stochastic parton scatterings, depends on the thermal medium density, which is characterized by the temperature dependence of the imaginary potential $V_I$. Previous theoretical studies suggest that the magnitude of the imaginary potential, $|V_I|$, increases linearly with temperature ($|V_I| \propto T$) and grows with the dipole separation according to a power-law form ($|V_I| \propto r^n$). Building upon these insights, we adopt the following parametrization for $V_I$~\cite{Zheng:2025bdc},
\begin{equation}
\label{eq:14}
V_I = - i T^{a_{0}} \left( a_1 r + a_2 r^{a_3} \right),
\end{equation}
where $i$ denotes the imaginary unit. The parameter $a_0$ characterizes the temperature dependence, while $a_1$, $a_2$, and $a_3$ describe the spatial dependence. In Eq.~(\ref{eq:14}), the variables $T$ and $r$ are dimensionless quantities representing the temperature and distance measured in units of GeV and fm, respectively. $V_I$ has the dimensions of energy (GeV). The specific values of the parameters $(a_0, a_1, a_2, a_3)$ will be extracted from experimental bottomonium data through deep learning methodologies.

For the real part of the heavy quark potential, the color screening effect from thermal partons can be encoded via the Debye mass $\mu_D(T)$~\cite{Lafferty_2019}: 
\begin{align}
    \label{eq:13}
V_R(T,r) = -\alpha\!\left(\mu_D + \frac{e^{-\mu_D r}}{r}\right)
           + \frac{2\sigma}{\mu_D}
           - \frac{\sigma e^{-\mu_D r}\,(2 + \mu_D r)}{\mu_D}.
\end{align}
and the Debye mass $\mu_D(T)$ is parametrized as,  
\begin{align}
    \label{eq:15}
\mu_D &= a_4 T \sqrt{4\pi N_c \left( 1 + \frac{N_f}{6} \right) \frac{\alpha}{3}}.
\end{align}
where $N_c = 3$ and $N_f = 3$ denote the number of colors and quark flavors in the bulk medium, respectively. The parameter $a_4$ is introduced to modulate the magnitude of the Debye mass, a smaller value of $a_4$ implies a weakened thermal medium effect on the real component of the heavy quark potential. As the temperature decreases, the in-medium heavy quark potential $V(T, r)$ is expected to evolve toward the vacuum Cornell potential $V_c$. To allow for a more generalized parametrization, we incorporate an additional parameter, the ``switching temperature" $T_{\text{sw}}$. At this threshold, the medium-dependent potential $V(T, r)$ transitions to the vacuum Cornell potential $V_c(r)$. This parameter effectively defines a regime where the hot QCD medium effects on the heavy quark potential become negligible for $T < T_{\text{sw}}$. Consequently, the description of the in-medium potential is extended as follows:
\[
V(T,r) = 
\begin{cases}
V_R(T,r) +V_I(T,r), & T \geq T_{\rm sw} \\
V_c(r), & T < T_{\rm sw}
\end{cases}
\]
Notably, if the value of $T_{\text{sw}}$ extracted by the deep neural networks approaches the pseudo-critical temperature $T_c$, it would imply that the current parametrizations of $V_R$ and $V_I$ are sufficiently robust to describe the experimental bottomonium data, rendering an additional switching temperature unnecessary. Given the crossover nature of the phase transition, we set the pseudo-critical temperature at $T_c = 0.17$ GeV. This temperature serves as the boundary separating the purely deconfined medium (for $T > T_c$) from the hadron gas (for $T < T_c$). To facilitate the sampling of diverse heavy quark potential configurations for the Schr\"odinger equation, the designated parameter ranges for the $V(T, r) = V_R + V_I$ framework are summarized in Table~\ref{tab:sample-para}.
\begin{table}[h!]
\centering
\caption{The sampling ranges for the heavy quark potential parameters are defined as follows: $a_0$ characterizes the temperature dependence, while $a_1, a_2,$ and $a_3$ dictate the spatial structure of the imaginary potential $V_I(T, r)$. The parameter $a_4$ modulates the strength of the Debye mass within the real component $V_R(T, r)$. Additionally, $T_{\text{sw}}$ represents the switching temperature at which the parametrization transitions between the in-medium form ($V_R + V_I$) and the vacuum case.}
\label{tab:sample-para}
\begin{tabular}{cc}
\hline\hline
Parameter & Sampling Range \\ \hline
$a_0$                & $[1.0,\;2.0]$      \\
$a_1$                & $[0.0,\;0.5]$      \\
$a_2$                & $[0.2,\;0.7]$    \\
$a_3$                & $[2.0,\;3.0]$        \\
$a_4$                & $[0.1,\;1.0]$      \\
$T_{\text{sw}}$& $[0.17,\;0.25]$  \\ \hline
\end{tabular}
\end{table}

\begin{figure}[htbp]
  \centering
\includegraphics[width=0.45\linewidth]{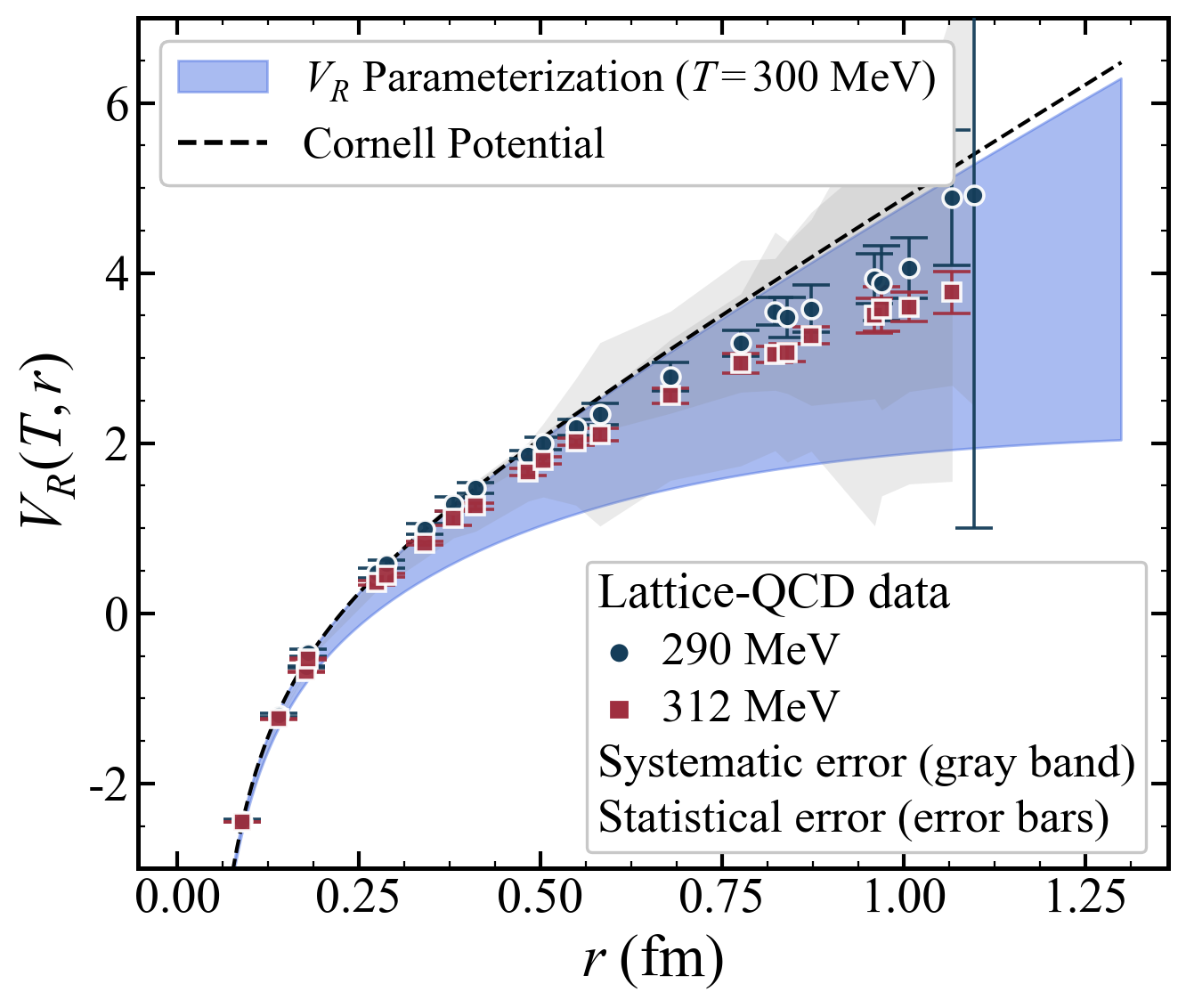}
\includegraphics[width=0.46\linewidth]{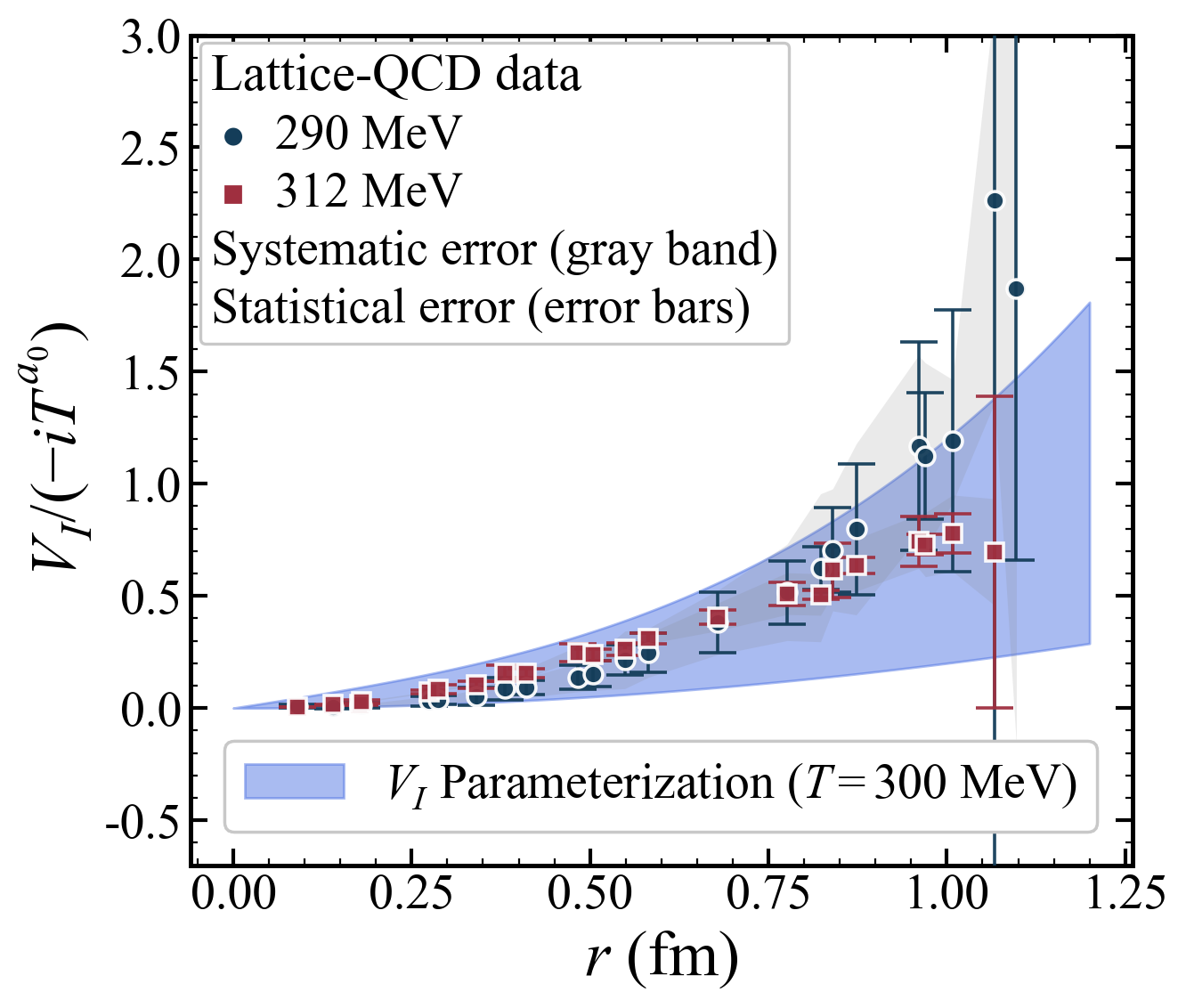}
\caption{Left panel: The real component of the heavy quark potential as a function of radius $r$ at a fixed temperature of $T = 300$~MeV. Right panel: The imaginary component of the potential, scaled by the temperature factor as $V_I / (-iT^{a_0})$, plotted versus $r$. In both panels, the shaded bands are composed of 1,000 individual curves, each representing a potential configuration sampled within the parameter ranges specified in Table~\ref{tab:sample-para}. The colored error bars in the figure represent the lattice QCD results at discrete points from Ref.~\cite{Burnier_2017}.
  }
\label{fig:pot-both}
\end{figure}

Utilizing various samples of the potential parameters listed in Table~\ref{tab:sample-para}, the corresponding real and imaginary components of the potentials are displayed as shaded bands in Figure ~\ref{fig:pot-both}. Each band in the subfigures comprises a collection of curves representing the parametrized real and imaginary potentials, respectively. In the panel for $V_R$, the vacuum Cornell potential ($V_c$) is included as a black thin dotted line, to provide a baseline for comparison. Figure~\ref{fig:pot-both} illustrates the broad range of complex potentials that serve as inputs for the time-dependent Schr\"odinger equation to generate the bottomonium $R_{AA}$ datasets. These datasets, which pair various heavy quark potential configurations with their corresponding $R_{AA}$ values, will be employed to train the deep neural networks.

\subsection{Hydrodynamic evolutions}

The hot deconfined medium generated in relativistic heavy-ion collisions has been found to behave as a nearly perfect fluid, which can be effectively described by relativistic hydrodynamic models. In this work, we employ the MUSIC package~\cite{schenke2011elliptic,schenke20103+1d} to generate the space-time temperature profiles, $T(\mathbf{x}, t)$, of the bulk medium, which subsequently serve as the thermal background for the in-medium heavy quark potentials $V(T, r)$. To close the hydrodynamic equations, we adopt an equation of state (EoS) that incorporates a crossover phase transition; specifically, the EoS for the deconfined and hadronic phases is based on parametrizations from lattice QCD calculations and the Hadron Resonance Gas model, respectively~\cite{Bernhard:2016tnd,HotQCD:2014kol}. Regarding the initial conditions for the hydrodynamic evolution, the Glauber model~\cite{loizides2016glauber} is utilized to determine the initial energy density distributions. The medium is assumed to reach local thermal equilibrium at a timescale of $t_0$, after which its dynamical expansion is governed by the hydrodynamic equations. The initialization time is set to $t_0 = 0.6$~fm/c~\cite{Shen:2014vra}, a value determined by fitting the final-state spectra of light hadrons in nuclear collisions. The maximum initial temperatures for the most central collisions (at impact parameter $b = 0$) are summarized in Table~\ref{tab:init-temp}, these values are constrained by the experimentally measured charged-particle multiplicity of light hadrons.

\begin{table}[htbp]
\centering
\caption{Initial conditions at midrapidity in different collision systems.}
\label{tab:init-temp}
\begin{tabular}{ccc}
\hline\hline
Collision System & \(t_0\;(\mathrm{fm}/c)\) & \(T_0(x_T{=}0|b=0)\;(\mathrm{MeV})\) \\ \hline
\(\sqrt{s_{_{NN}}}=5.02\;\text{TeV (Pb--Pb)}\) & 0.6 & 510 \\
\(\sqrt{s_{_{NN}}}=2.76\;\text{TeV (Pb--Pb)}\) & 0.6 & 484 \\
\(\sqrt{s_{_{NN}}}=200\;\text{GeV (Au--Au)}\) & 0.6 & 390 \\ \hline\hline
\end{tabular}
\end{table}

\begin{figure}[htbp]%[htbp]
  \centering
\includegraphics[width=0.94\linewidth]{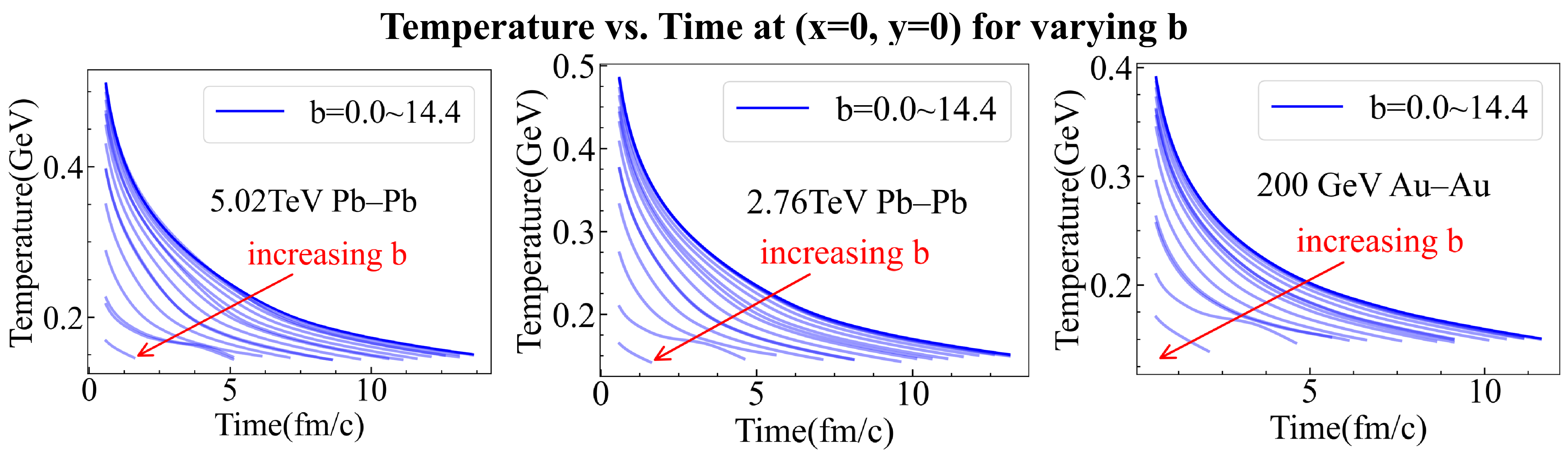}
  \caption{Temporal evolution of the local temperature at the center of the fireball for $5.02$~TeV Pb-Pb, $2.76$~TeV Pb-Pb, and $200$~GeV Au-Au collisions. Each colored curve represents a specific impact parameter $b$, varying from $0.0$ to $14.4$~fm in increments of $1.2$~fm.}
  \label{fig:temp-t}
\end{figure}

Based on the MUSIC package, the temporal evolution of the temperature at the center of the fireball, $T(\mathbf{x}_T = 0, t)$, for various collision centralities is shown in Figure ~\ref{fig:temp-t}. The three panels correspond to the three collision systems: $\sqrt{s_{NN}} = 5.02$~TeV Pb-Pb, $2.76$~TeV Pb-Pb, and $200$~GeV Au-Au collisions. In each panel, the distinct curves represent the temperature evolution, $T(\mathbf{x}_T = 0, t)$, for different impact parameters ($b$).

\subsection{Training dataset generation}

The in-medium heavy quark potential is parametrized by a six-dimensional vector: $\mathbf{x}_V = (a_0, a_1, a_2, a_3, a_4, T_{\text{sw}})$. These parametrized potentials are incorporated into the Schr\"odinger equation framework to calculate the bottomonium nuclear modification factors ($R_{AA}$) across various centrality and transverse momentum bins. Specifically, the theoretical curves for $R_{AA}(N_{part})$ and $R_{AA}(p_T)$ for each state, $\Upsilon(1S)$, $\Upsilon(2S)$, and $\Upsilon(3S)$, are represented by 13 and 9 data points, respectively. The output of the Schr\"odinger model, denoted as $\mathbf{y}_V$, yields a total of $3 \times (13 + 9) = 66$ data points for each potential sample $\mathbf{x}_V$. At $\sqrt{s_{NN}} = 2.76$~TeV Pb-Pb collisions, while the input structure remains $\mathbf{x}_V$, experimental measurements are available only for the $\Upsilon(1S)$ and $\Upsilon(2S)$ states; thus, the corresponding model output consists of $2 \times (13 + 9) = 44$ data points. Similarly, for $200$~GeV Au-Au collisions, where only the $\Upsilon(1S)$ and $\Upsilon(2S)$ states are measured, the output is reduced to $2 \times (13 + 7) = 40$ data points. In the subsequent analysis, we utilize the $5.02$~TeV Pb-Pb system as a representative case. We randomly generate $N=1000$ samples of the potential parameter vector ${\bf X_V}$, each of them incorporated into the Schr\"odinger equation model to calculate the corresponding bottomonium $R_{AA}$, denoted as ${\bf Y_V}$. To ensure evaluation performance, the $N=1000$ samples are split $8 : 2$ where
80\% of the events are used for training the following Principal Component Analysis~\cite{PCAanalysisref1,Bernhard:2015hxa} and the Gaussian Process emulator~\cite{Rasmussen2006Gaussian}, which helps to generate larger datasets used in the training of the deep neural networks, while the remaining
20\% of $N=1000$ events are reserved as an independent test set for the subsequent deep neural networks.
The resulting input and output datasets are structured as matrices $\mathbf{X}_V \in \mathbb{R}^{N \times 6}$ and $\mathbf{Y}_V \in \mathbb{R}^{N \times 66}$, respectively, as defined below:
\[
\mathbf{X_V} = \begin{bmatrix}
x_{1,1}   & \dots & x_{1,6} \\
\vdots &  \ddots & \vdots \\
x_{N,1}   & \dots & x_{N,6}
\end{bmatrix}, \qquad\\
\mathbf{Y_V} = \begin{bmatrix}
y_{1,1}  & \dots & y_{1,66}\\
\vdots &  \ddots & \vdots \\
y_{N,1}   & \dots & y_{N,66}
\end{bmatrix} 
\]

Although the output vector $\mathbf{y}_V$ comprises 66 $R_{AA}$ points, these variables are highly correlated. To address this, it is essential to project this high-dimensional, correlated output onto a lower-dimensional space using Principal Component Analysis. This dimensionality reduction captures the dominant features and variance within the datasets, thereby streamlining the training process for the deep neural network. 
Before applying PCA, the training output matrix ${\bf Y_V}$ is standardized by subtracting the mean and dividing by the standard deviation for each feature. 
To achieve dimensionality reduction, we retain the first $k$ principal components. The first $k = (4, 3, 3)$ principal components are sufficient to capture 99\% of the total variance across the respective collision systems. These dimension-reduced datasets will subsequently serve as inputs for the Gaussian Process emulator to generate the augmented datasets required for the deep neural network analysis.
\begin{figure}[htbp]
\centering
\includegraphics[width=0.5\linewidth]{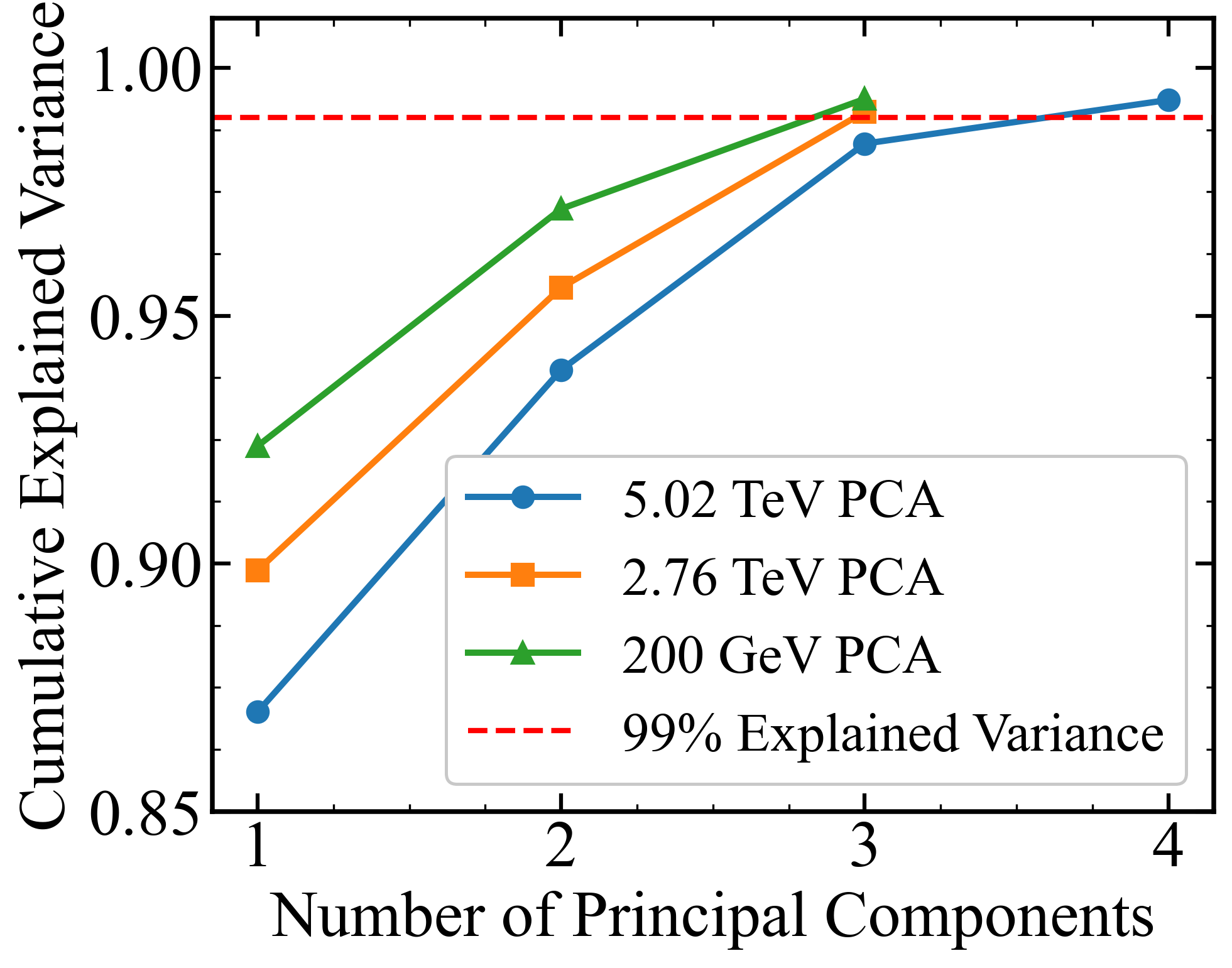}
  \caption{The cumulative explained variance of the top \( k \) principal components for the raw \( R_{AA} \) data from the three different collision energies. The number of principal components \( k \) is taken as \( k = (1, 2, 3, 4) \), respectively.}
  \label{fig:pca}
\end{figure}

In a Gaussian process emulator, 
$N_{\rm GP}$ events (comprising 80\% of the initial 1000 potential parameter sets) are utilized. To evaluate the predictive performance of the GP emulators, $N_{\rm GP}$ is further partitioned into training and testing subsets at a 9:1 ratio. Specifically, $90\%$ of $N_{\rm GP}$ is used for PCA and GP training, while the remaining $10\%$ ($N_{\rm test}$) serves as an independent test set to calculate the coefficient of determination, $R^2_j$. This metric represents the emulator accuracy for the $j$-th principal component,
\begin{align}
\label{eq-R2-eff}
R^2_j = 1- \frac{\sum_{i=1}^{N_{\rm test}} \bigl(z^{(j)}_i - \widehat{z}^{(j)}_i\bigr)^2}{\sum_{i=1}^{N_{\rm test}} \bigl(z^{(j)}_i - \bar{z}^{(j)}\bigr)^2},
\end{align}
where, $\hat{z}_i^{(j)}$ denotes the true value of the $i$-th event in the dataset, while $z_i^{(j)}$ represents the predicted output for the same event projected onto the $j$-th principal component, as provided by the GP emulators. Each event corresponds to a unique set of potential parameters. The term $\bar{z}^{(j)}$ refers to the mean value of the outputs projected onto the $j$-th principal component. Higher $R^2_j$ values signify superior emulator accuracy within the potential parameter space. The $R^2_j$ scores for each principal component (evaluated on the test set) are all above 0.96 across the three collision energies. These high scores demonstrate that the GP emulators are effectively trained to capture the essential features of the dataset while maintaining robust generalization and avoiding overfitting.

\begin{figure}[htbp]
  \centering
  \begin{minipage}[t]{0.45\linewidth}
    \centering
    \includegraphics[width=\linewidth]{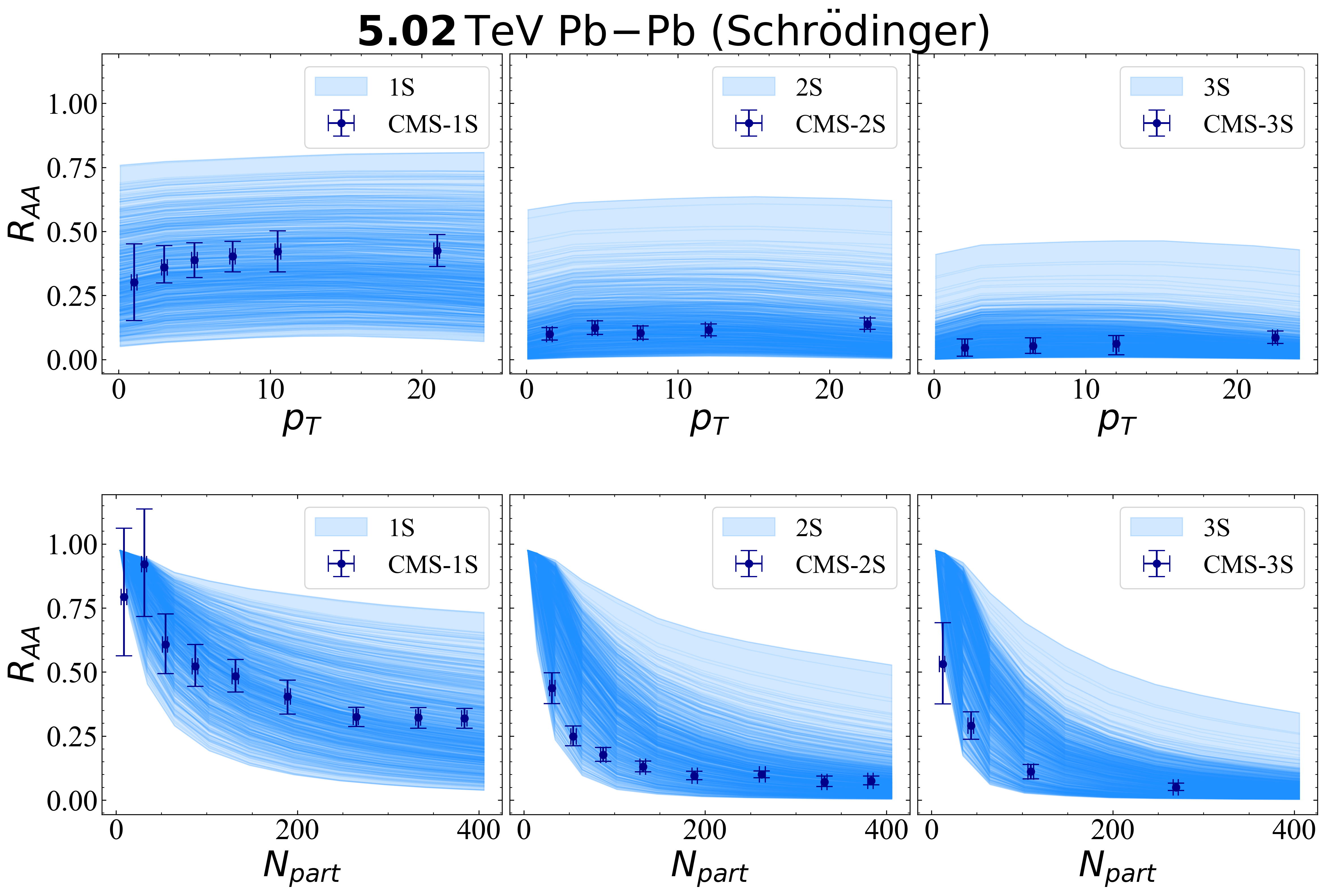}
  \end{minipage}%
  \hfill%
  \begin{minipage}[t]{0.45\linewidth}
    \centering
    \includegraphics[width=\linewidth]{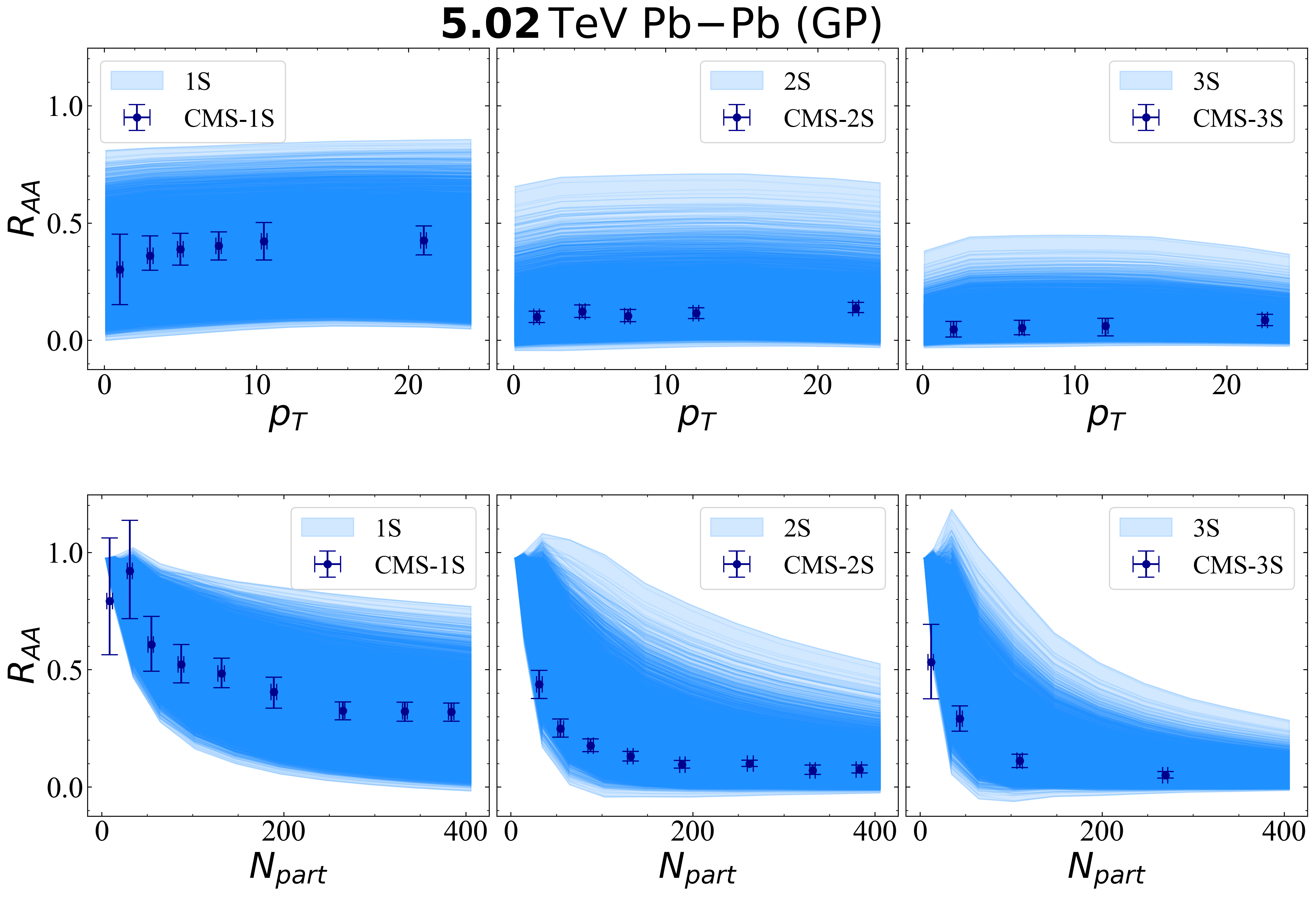}
  \end{minipage}
  \caption{Left panel: the bands consist of $1,000$ events from Schr\"odinger model calculations, which span the ranges of experimental data for \( R_{AA} \) as functions of \( N_{\rm part} \) and \( p_T \) at 5.02 TeV. Right panel: The bands consist of $10,000$ events generated by the Gaussian process emulator, compared with experimental data at the same collision energy. }
  \label{fig:three-sch-gaussi}
\end{figure}

\begin{figure}[!h]
  \centering
  \begin{minipage}[t]{0.45\linewidth}
    \centering
    \includegraphics[width=\linewidth]{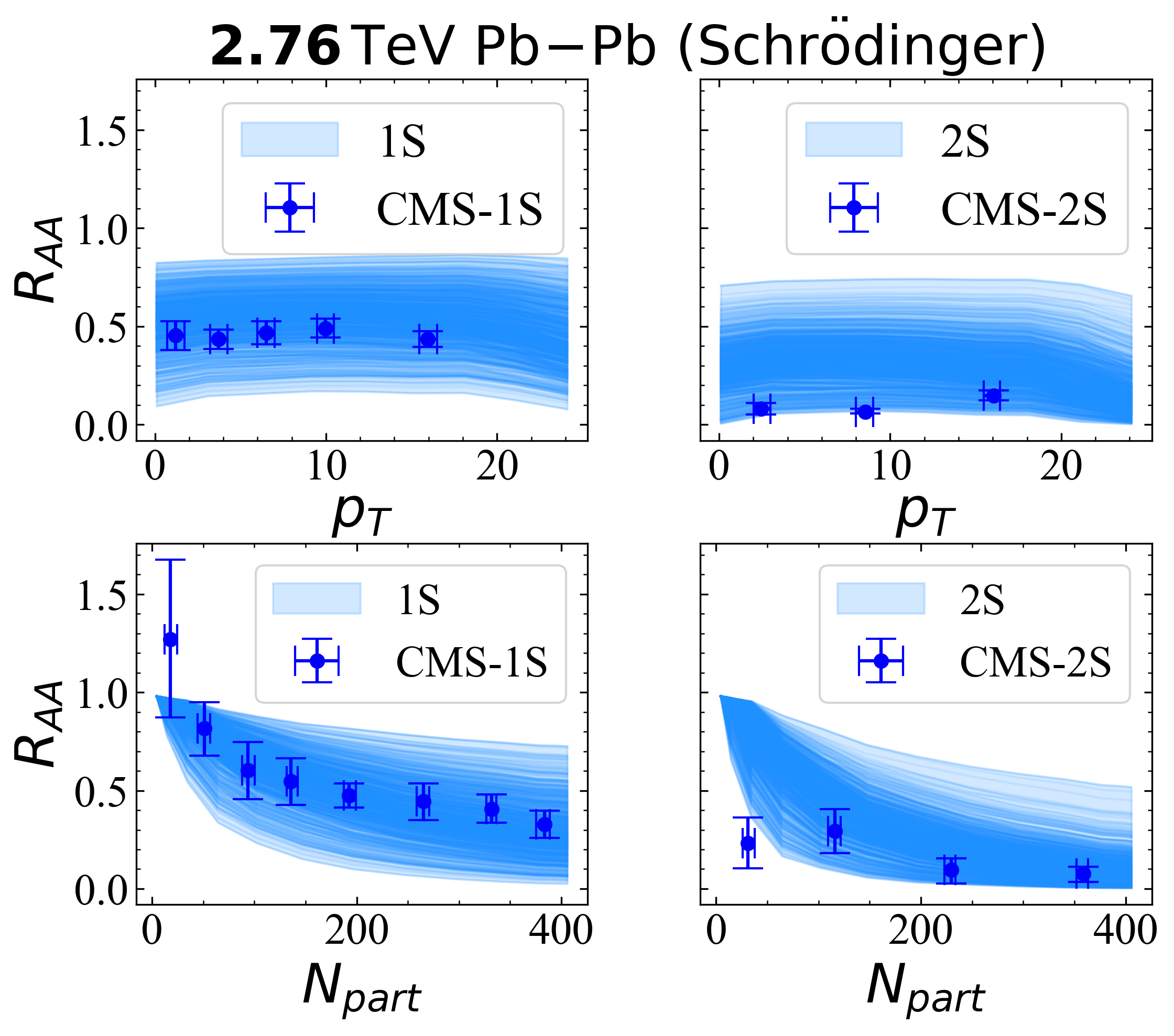}
  \end{minipage}%
  \hfill%
  \begin{minipage}[t]{0.45\linewidth}
    \centering
    \includegraphics[width=\linewidth]{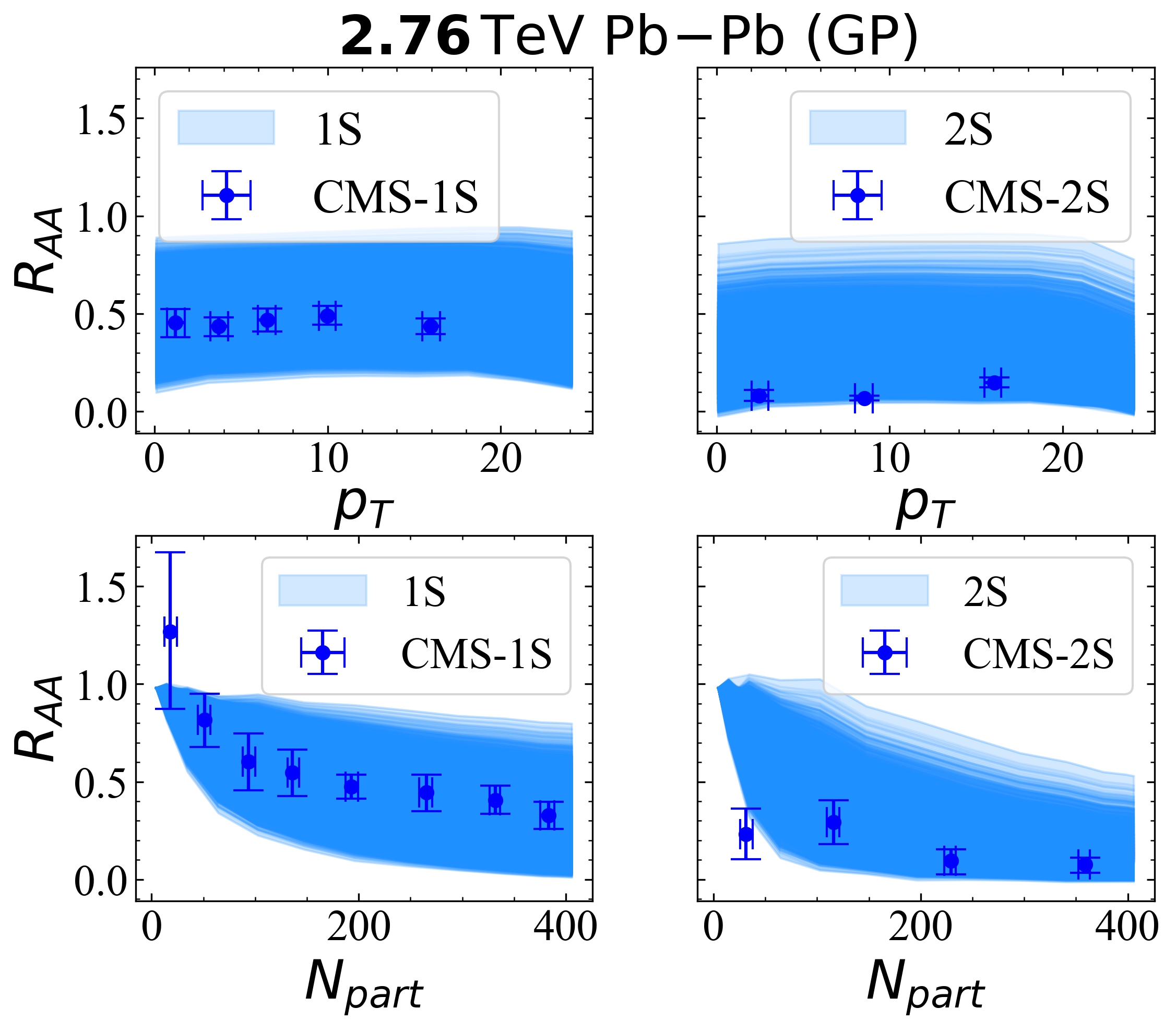}
  \end{minipage}
  \caption{Same as Figure \ref{fig:three-sch-gaussi}, but presented specifically for the $\sqrt{s_{NN}} = 2.76$ TeV Pb-Pb collisions. }
  \label{fig:2.76-sch-gaussi}
\end{figure}

\begin{figure}[htbp]
  \centering
  \begin{minipage}[t]{0.45\linewidth}
    \centering
    \includegraphics[width=\linewidth]{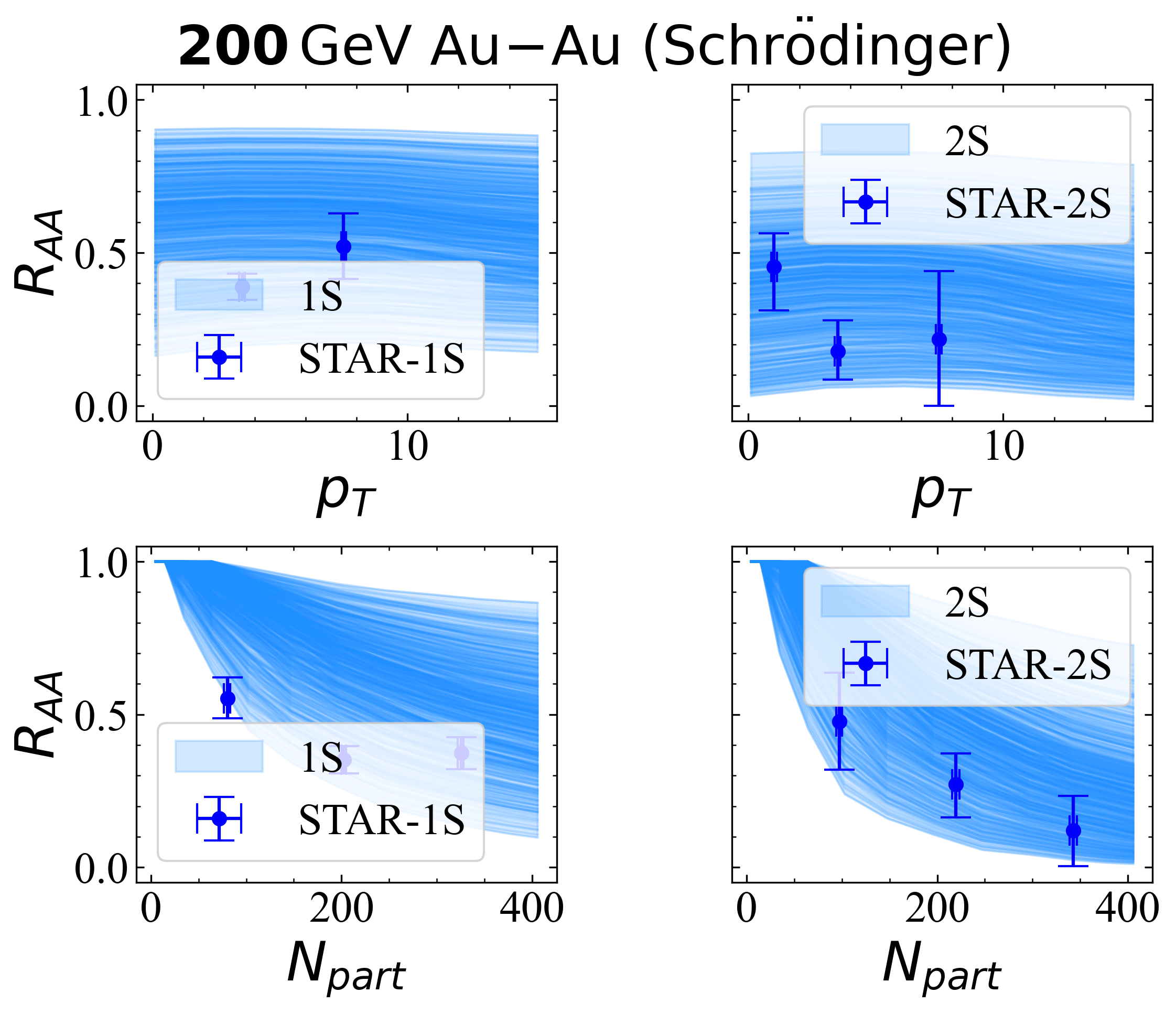}
  \end{minipage}%
  \hfill%
  \begin{minipage}[t]{0.45\linewidth}
    \centering
    \includegraphics[width=\linewidth]{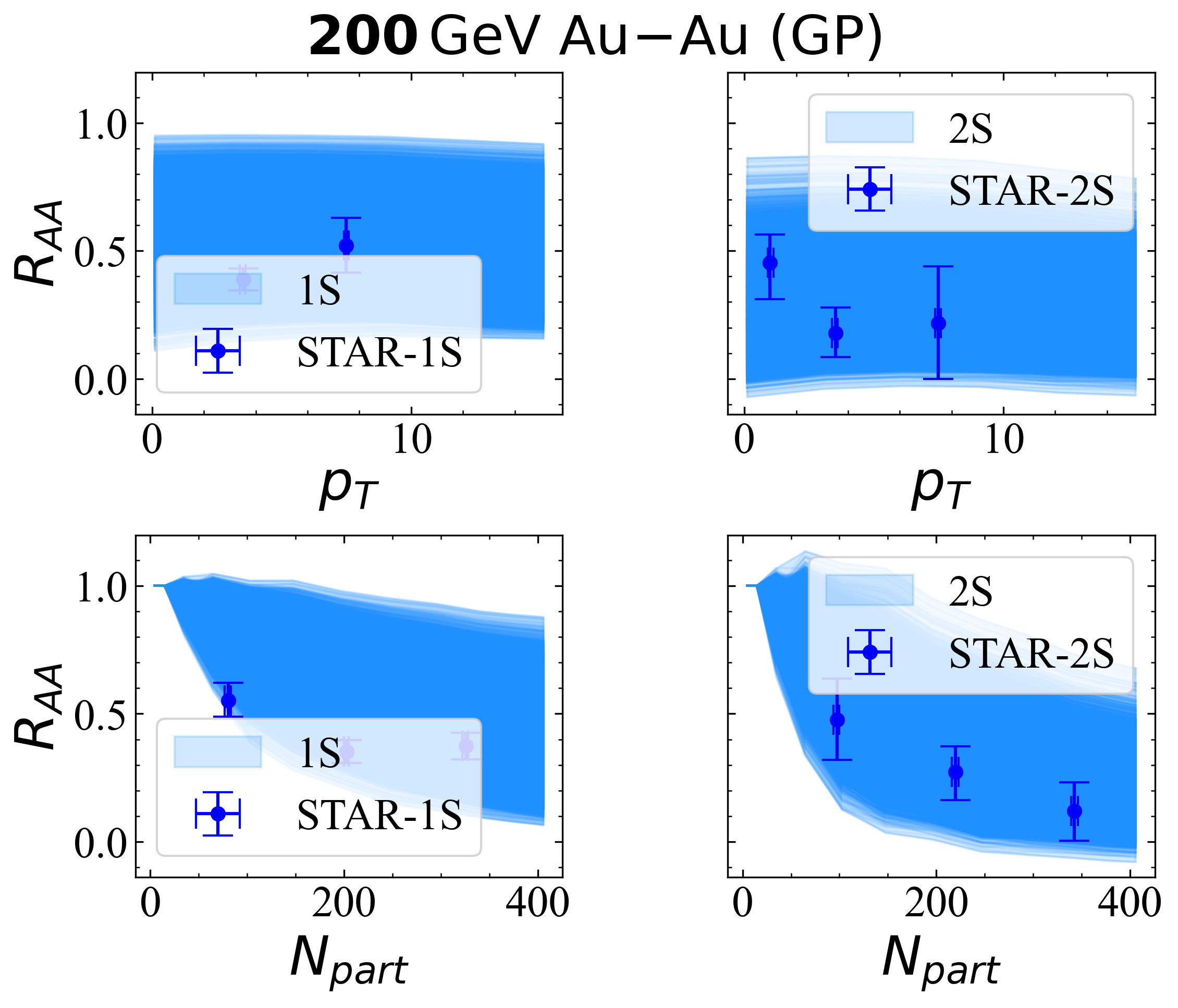}
  \end{minipage}
  \caption{Same as Figure \ref{fig:three-sch-gaussi}, but presented specifically for the $\sqrt{s_{NN}} = 200$ GeV Au-Au collisions. }
  \label{fig:200-sch-gaussi}
\end{figure}

By extending the limited theoretical dataset derived from the Schrödinger model, this strategy significantly expands data coverage and enhances the DNN’s ability to capture the underlying patterns of bottomonium $R_{AA}$ across varying collision energies. The left and right panels of Figure  \ref{fig:three-sch-gaussi} present the events generated by the Schrödinger model with those predicted by the GP emulator. These comparisons validate our simulation framework and ensure that the dataset provided to the DNN models adequately spans the range of experimental observations. The corresponding results for 2.76 TeV Pb-Pb and 200 GeV Au-Au collisions are presented in Figs. \ref{fig:2.76-sch-gaussi} and \ref{fig:200-sch-gaussi}, respectively.

\section{Deep learning applications in heavy-ion collisions}
\label{sec:train-cnn}

\subsection{Deep neural network}

\begin{figure}[htbp]
\centering    
\includegraphics[width=0.85\textwidth]{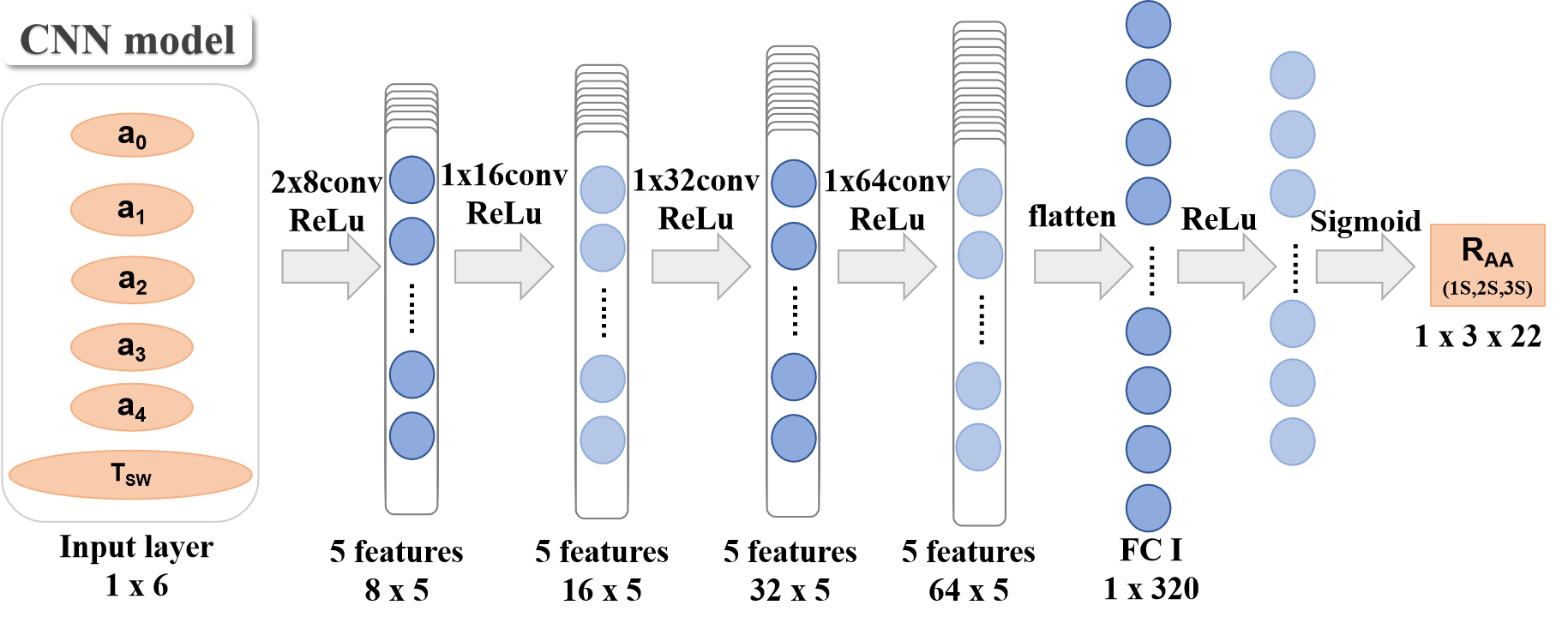} 
\caption{Architecture of the Convolutional Neural Network in this work.}
\label{fig:cnn}
\end{figure}

We construct independent Convolutional Neural Networks for each collision system, including 5.02 TeV Pb-Pb, 2.76 TeV Pb-Pb, and 200 GeV Au-Au. Taking the 5.02 TeV Pb-Pb system as a representative example, the CNN architecture is illustrated in Figure  \ref{fig:cnn}, comprising input, hidden, and output layers. The CNN input consists of potential parameters $(a_0, a_1, a_2, a_3, a_4, T_{\rm sw})$, which are randomly sampled within the ranges specified in Table \ref{tab:sample-para}. The output is a vector of bottomonium $R_{AA}$ values across various centrality and transverse-momentum bins. To introduce necessary nonlinearity into the model, the Rectified Linear Unit (ReLU) is employed as the activation function for the hidden layers.

The CNN architecture processes a 3D input tensor with dimensions (batch size, channels, features). For the training phase, we employ a mini-batch size of 64, yielding an input shape $[B,1,6]$ with $B=64$, where the second dimension represents a single channel. Following the convolutional blocks, the resulting tensor is flattened and passed through two fully connected (FC) layers. The first FC layer transforms the feature vector from a shape of $[B, 320]$ to $[B, 128]$, while the second layer maps it to $[B, 66]$. Finally, the output is reshaped to $[B, 3, 22]$, corresponding to the predicted $R_{AA}(N_{\mathrm{part}})$ and $R_{AA}(p_T)$ values for three bottomonium states: $\Upsilon(1S)$, $\Upsilon(2S)$, and $\Upsilon(3S)$. Here, the value of 22 denotes the total number of data points sampled from the theoretical $R_{AA}$ curves.

In the CNN architecture, the Adam optimizer is employed to minimize the loss function. To prevent overfitting within the neural network, an $L_2$ regularization term is incorporated into the loss function.

\subsection{Bayesian Inference and SGLD sampling}

For each collision system, an independent CNN is constructed and trained. The model establishes a mapping from the in-medium heavy quark potentials, parameterized by $\boldsymbol{\theta}=(a_0,a_1,a_2,a_3,a_4,T_{\rm sw})$, to the bottomonium $R_{AA}$ observables. The output of the CNN is denoted as $\mathbf{y}_{\text{CNN}}(\boldsymbol{\theta})$. Experimentally, the nuclear modification factors for different bottomonium states have been measured across various centralities and are denoted by $\mathbf{y}_{\text{obs}} = \{y_{\text{obs},i}\}$. Each data point $y_{\text{obs},i}$ is associated with an experimental uncertainty $\sigma_i$, which determines its respective weight in the extraction of the in-medium potential.

In Bayesian analysis to extract the values of the potential parameters, the corresponding chi-squared $\chi^2$ is defined as, 
\begin{align}
\chi^2(\boldsymbol{\theta})
=\sum_{i=1}^{N_e} \frac{\bigl[y_{\text{obs},i}-y_{{\rm CNN},i}(\boldsymbol{\theta})\bigr]^2}{\sigma_i^2},
\label{eq:chi2-diag}
\end{align}
where $N_e$ denotes the total number of experimental data points. 
Calculating the $\chi^2$ statistic requires consistency between the experimental data points and the CNN outputs at identical abscissae, specifically, the $N_p$ and $p_T$ coordinates. However, while the CNN outputs are generated on fixed grids (e.g., 13 bins in $N_p$ and 6 bins in $p_T$), the abscissae $x_i$ of the experimental data points often do not align with these grid positions. To resolve this discrepancy, we define a linear interpolation operator $\mathcal{I}_g$ for each data subset $g$. This operator maps the gridded CNN outputs to a new vector whose abscissae precisely match those of the experimental data points.
\begin{equation}
\mathbf{y}^{(g)}_{{\rm CNN}\to{\rm obs}}(\boldsymbol{\theta})
=\mathcal{I}_g\,\mathbf{y}_{\rm CNN}(\boldsymbol{\theta}).
\label{eq:interp}
\end{equation}
Consequently, the interpolated vector $\mathbf{y}^{(g)}_{\text{CNN}\to\text{obs}}(\boldsymbol{\theta})$ maintains the same dimensionality as the experimental observation vector $\mathbf{y}_{\text{obs}}$. 
For both optimization and Stochastic Gradient Langevin Dynamics (SGLD) updates, the data-dependent objective can be defined as the negative log-likelihood,
\begin{equation}
\mathcal{L}(\boldsymbol{\theta})
= -\log p(\mathbf{y}_{\rm obs}\mid\boldsymbol{\theta})
= \sum_{g}\ \sum_{i\in g}
\frac{1}{2}\,
\frac{\Bigl(\mathbf{y}^{(g)}_{{\rm CNN}\to{\rm obs}, i}(\boldsymbol{\theta})
-\mathbf{y}^{(g)}_{{\rm obs}, i}\Bigr)^2}
{\sigma_i^2}.
\label{eq:weighted-loss}
\end{equation}

For the prior distribution of the potential parameters, 
we assume a uniform prior distribution supported on the physically admissible domain $\Omega$. 
Rather than performing numerical sampling for $\boldsymbol{\theta} \in \Omega$ directly on the bounded domain, where boundary discontinuities may adversely affect the sampling efficiency, we introduce a reparameterization $\boldsymbol{\theta} = \boldsymbol{\theta}(\boldsymbol{z}) \in \Omega$ as defined in Eq. (\ref{eq:sigmoid-map}). This transformation allows us to operate in an unconstrained auxiliary space, $\boldsymbol{z} \in \mathbb{R}^6$,
\begin{equation}
\boldsymbol{\theta}(\boldsymbol{z})
={\boldsymbol{\theta}}_{\rm min}
+\bigl({\boldsymbol{\theta}}_{\rm max}-{\boldsymbol{\theta}}_{\rm min}\bigr)
\odot \sigma(\boldsymbol{z}),
\label{eq:sigmoid-map}
\end{equation}
where $\sigma(z)=\frac{1}{1+e^{-z}}$ and \(\odot\) denotes the Hadamard product.

The parameters $\boldsymbol{z}$ and the resultant mappings $\boldsymbol{\theta(z)}$ are iteratively updated via the Stochastic Gradient Langevin Dynamics (SGLD) algorithm. In the iteration of the parameters $\boldsymbol{\theta(z)}$, a stochastic term sampled from a standard normal distribution at each training iteration is introduced to increase the stability of the SGLD process. 
The learning rate, $\eta_t$, follows a time-dependent schedule as detailed in the subsequent section. Given the functional dependence $\boldsymbol{\theta} = \boldsymbol{\theta}(\boldsymbol{z})$, the parameters $\boldsymbol{\theta}$ of in-medium heavy quark potentials are updated implicitly through the SGLD trajectories of $\boldsymbol{z}$ in the auxiliary space.

The uncertainty of the potential parameters is quantified using posterior samples generated directly via SGLD. We then determine the joint maximum a posteriori (MAP) estimates for the heavy quark potential parameters based on bottomonium experimental data ${\bf y}_{\rm obs}$,
\begin{equation}
\widehat{\boldsymbol{\theta}}_{\rm MAP}
=\arg\min_{\boldsymbol{\theta}\in\Omega}\,\mathcal{L}(\boldsymbol{\theta}).
\end{equation}
With the retained sample $\mathcal{S}_\gamma$ that minimizes the loss function $\mathcal{L}$ of DNN, one can obtain different values $\boldsymbol{\theta}\in \mathcal{S}_\gamma$ of the heavy quark potential parameters. These  samples will be used to extract the uncertainty of the heavy quark potential. A detailed quantification of the uncertainties associated with the heavy quark potential is provided below.

\begin{figure}[htbp]
\centering
\includegraphics[width=0.7\linewidth]{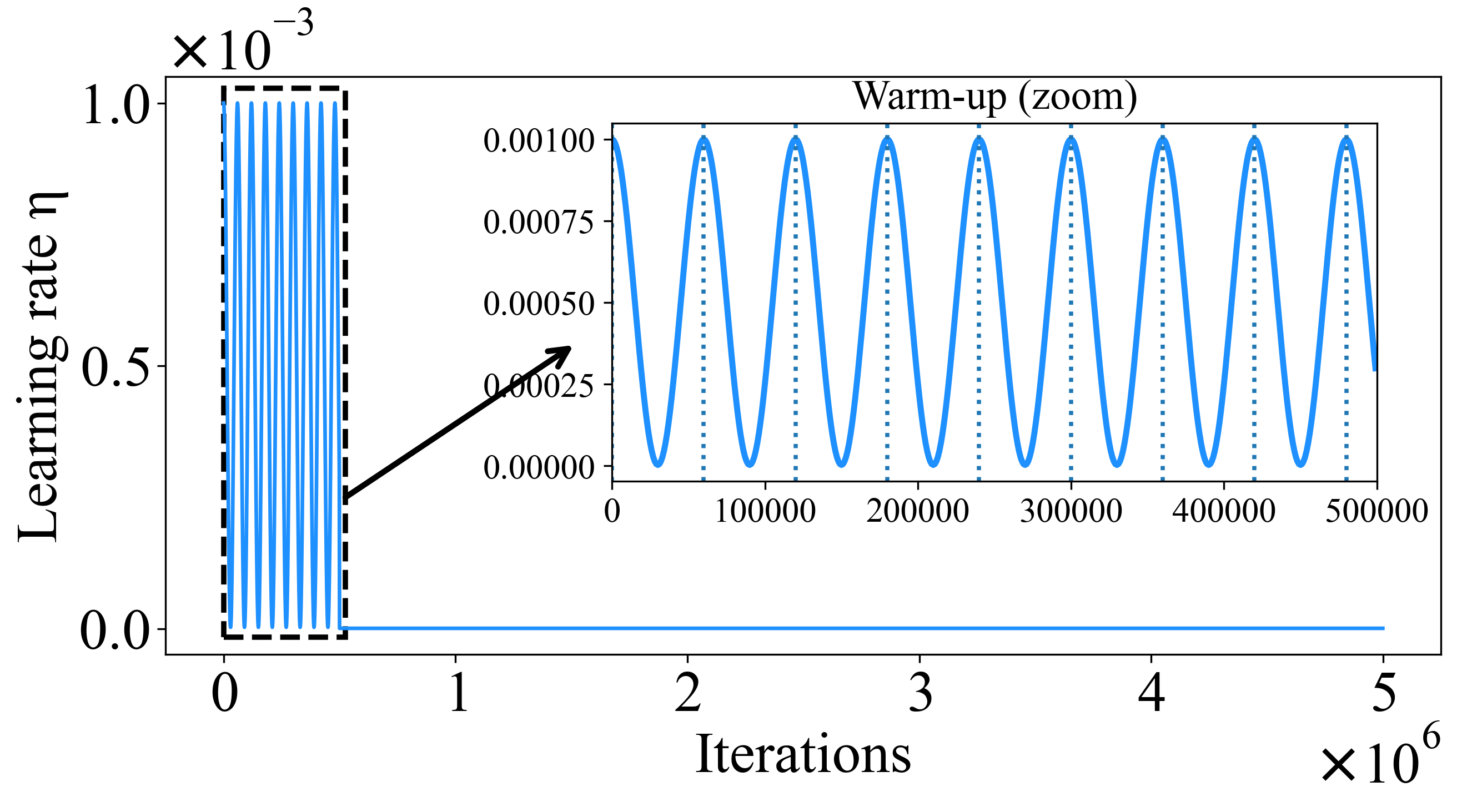}
\caption{The variable learning rate $\eta_t$ used in SGLD sampling is shown as a function of the iteration index $t$.}
\label{fig:lr}
\end{figure}

To enhance both the convergence speed and stability of SGLD, a variable learning rate schedule is employed. During the initial $N_{\rm warm}=5\times10^{5}$ iterations, the learning rate follows a cosine warm-up cycle before settling at a small constant value. This approach allows the model to progressively move into more favorable regions of the parameter space while mitigating excessive oscillations during parameter updates. For iterations $t < N_{\rm warm}$, the learning rate $\eta$ is cycled between $\eta_{\min}=3\times10^{-6}$ and $\eta_{\max}=10^{-3}$ with a period of $P=6\times10^{3}$ iterations. Subsequently, it is fixed at $\eta = \eta_{\text{const}} = 1\times10^{-6}$ for the remaining steps, as shown in Eq.~\eqref{eq:lr-coswarm-const} and Figure ~\ref{fig:lr},
\begin{equation}
\eta_t=
\begin{cases}
\displaystyle \eta_{\min}
+\frac{\eta_{\max}-\eta_{\min}}{2}\!\left(1+\cos\!\left(2\pi\,\frac{t \bmod P}{P}\right)\right),
& 0 \le t < N_{\mathrm{warm}},\\[8pt]
\eta_{\mathrm{const}}, & t \ge N_{\mathrm{warm}} .
\end{cases}
\label{eq:lr-coswarm-const}
\end{equation}
where $t\in\mathbb{N}$ denotes the iteration index. The parameters $\eta_{\min}$ and $\eta_{\max}$ represent the lower and upper bounds of the cosine schedule during the warm-up phase ($0\le t<N_{\rm warm}$) of the SGLD process.

\subsection{Performance of Deep Neural Networks}
After the training process, we obtain three distinct CNN models, each representing the mapping between the heavy quark potential and bottomonium $R_{AA}$ for three specific collision systems. In the subsequent heavy quark extraction phase, we employ an SGLD-based approach to incorporate experimental bottomonium data. By minimizing the sum of the loss functions from three well-trained CNN models, we identify a unified set of potential parameters. This global optimization ensures that the resulting potential parameters can simultaneously account for the bottomonium experimental data across all three collision energies. Prior to performing the SGLD sampling, we validate each trained CNN to verify its accuracy in predicting the $R_{AA}$ values produced by varying potentials.

\begin{figure}[htbp]
\centering
\includegraphics[width=0.3\linewidth]{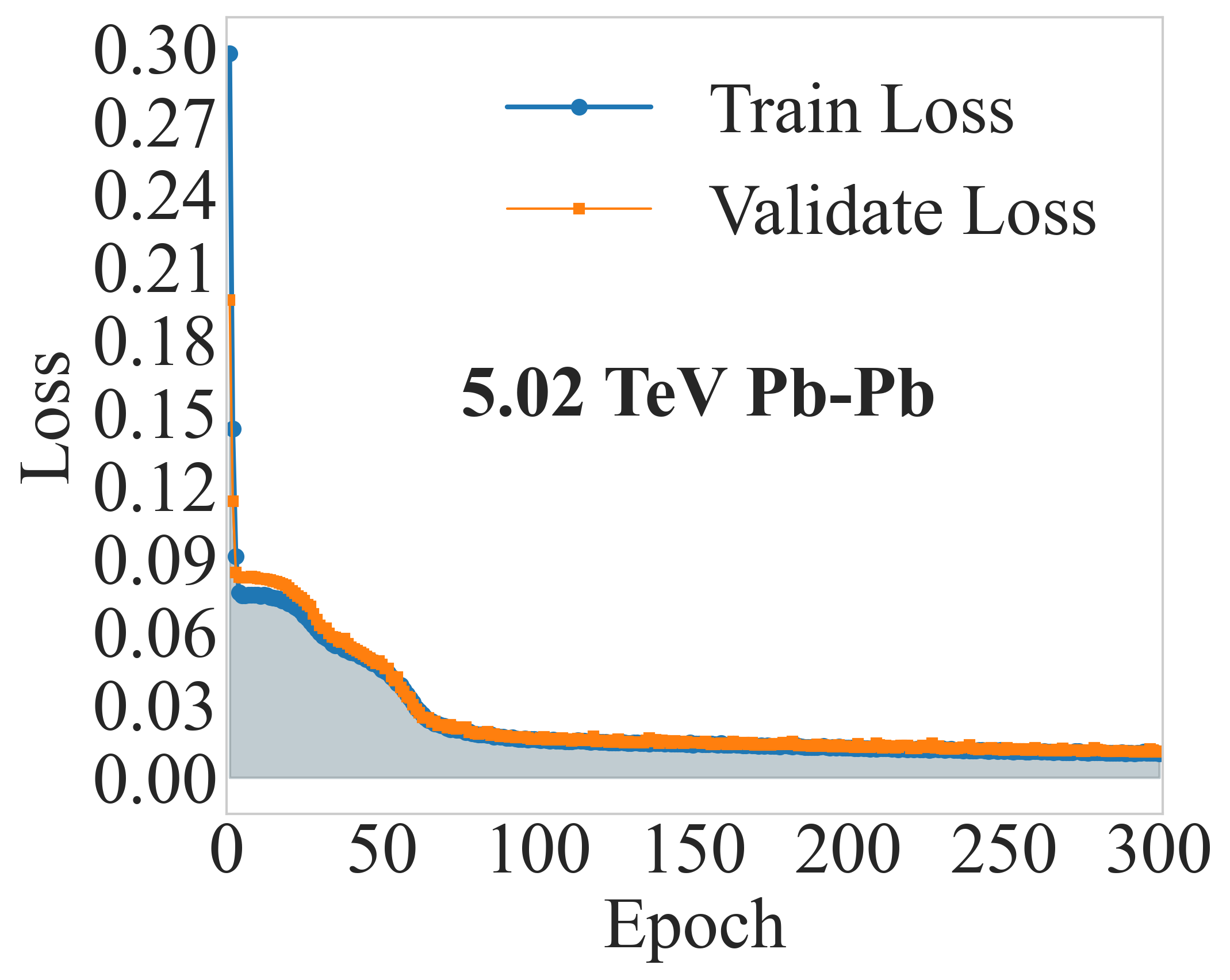}
\includegraphics[width=0.3\linewidth]{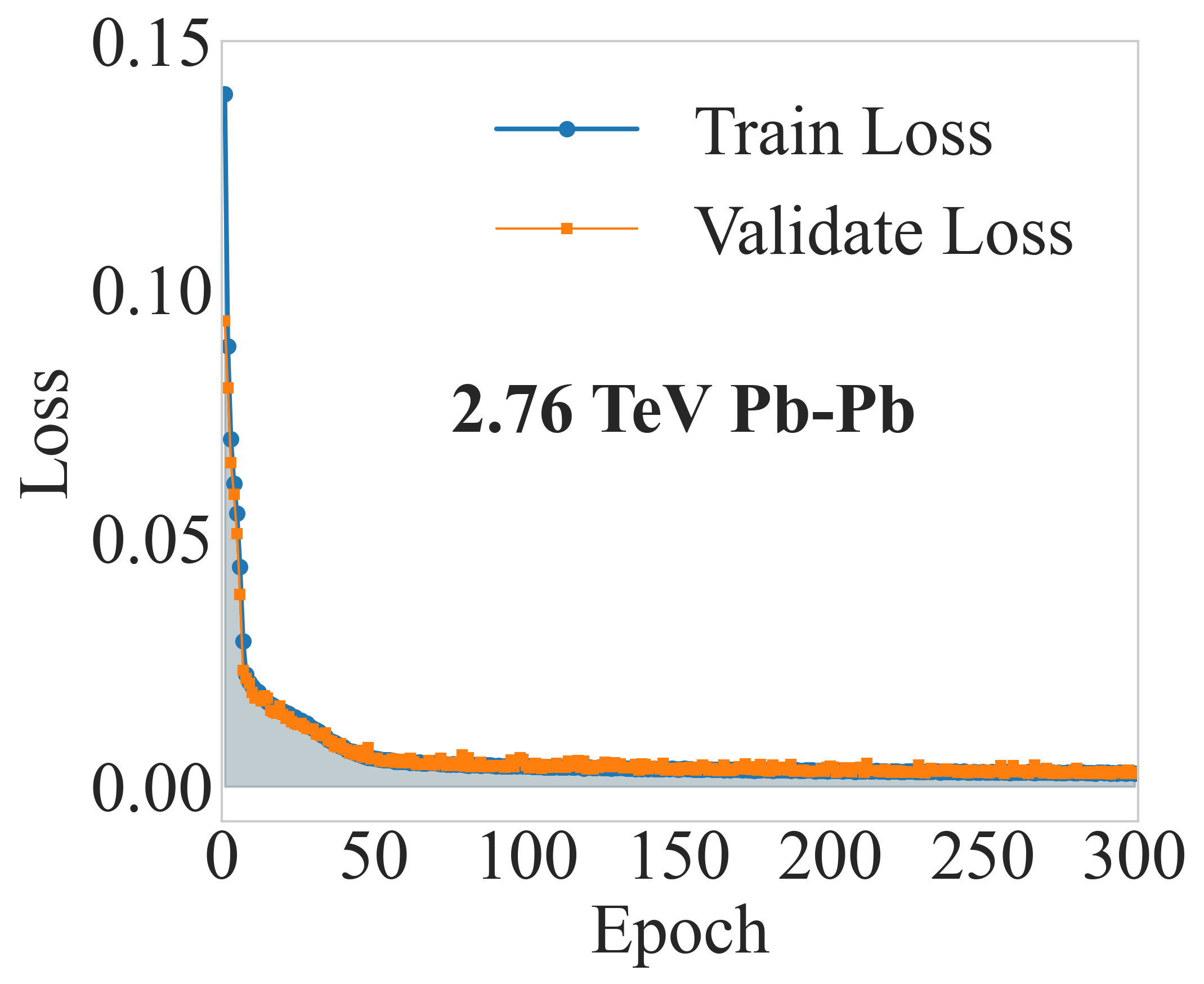}
\includegraphics[width=0.3\linewidth]{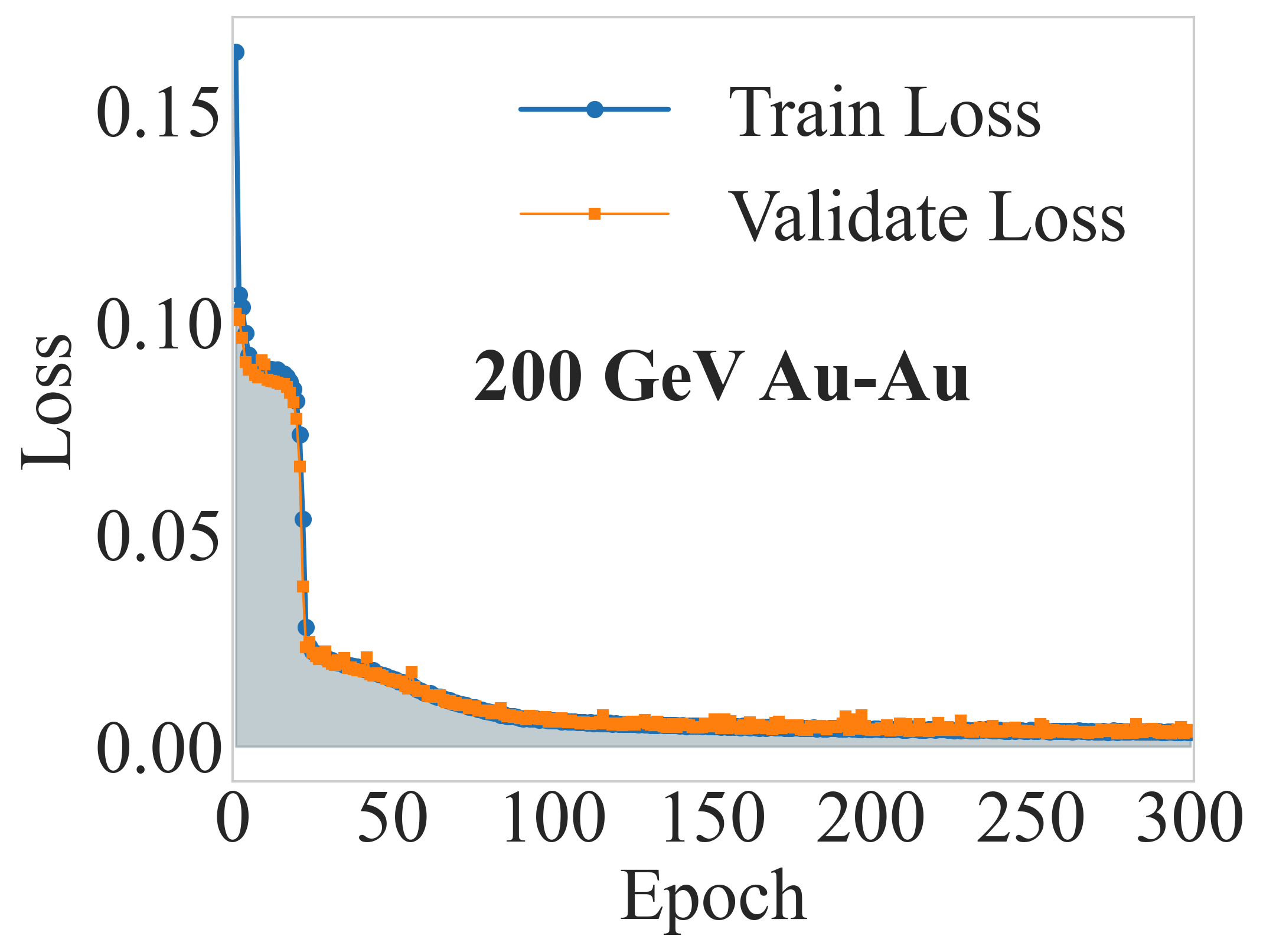}
  \caption{ Training loss as a function of the number of epochs for the three CNN models corresponding to different collision systems: 5.02 TeV Pb-Pb (left), 2.76 TeV Pb-Pb (middle), and 200 GeV Au-Au (right). Each model was trained on 10,000 samples. In all cases, the final loss values converge to below 0.01.}
  \label{fig:loss-3snn}
\end{figure}

With 80\% of $N=1,000$ potential parameter samples generated based on the Schr\"odinger model, we utilize the Principal Component Analysis (PCA) and Gaussian Process (GP) emulators to generate a larger dataset of 10,000 events for training the CNN models. 
After sufficient training iterations, the loss for each CNN model decreases to a value below 0.01 and plateaus, as shown in Figure ~\ref{fig:loss-3snn}, indicating that the networks have effectively captured the features of the bottomonium datasets for their respective collision systems. Our tests demonstrate that employing three specialized CNNs, one for each collision system, more effectively captures dataset-specific features compared to a single unified CNN trained on the combined data.

\begin{figure}[htbp]
\centering
\includegraphics[width=0.85\linewidth]{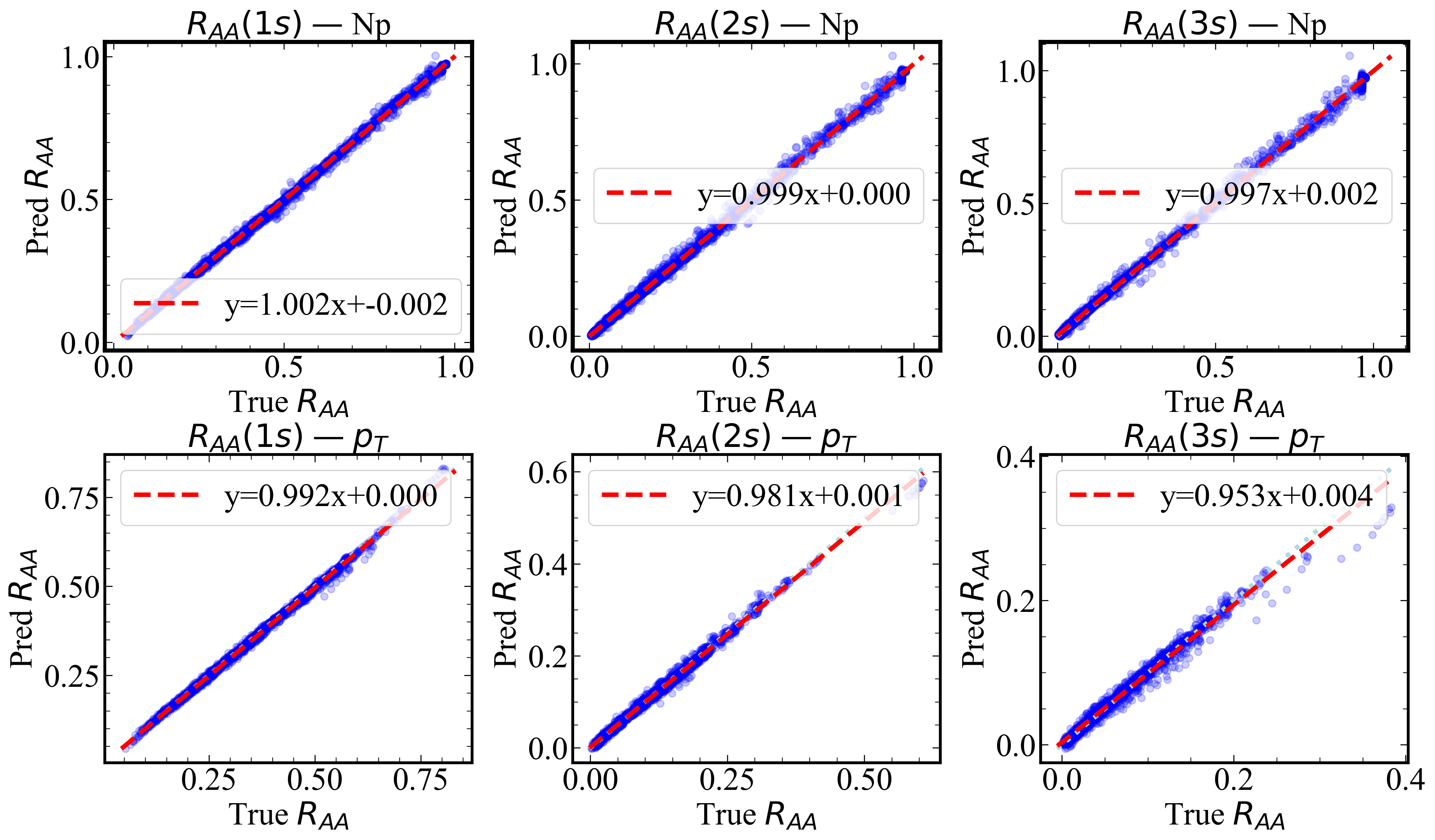}
  \caption{
Performance validation of the CNN model for 5.02 TeV Pb-Pb collisions. Using the trained CNN, we predict the bottomonium $R_{AA}$ for $0.2N$ heavy quark potentials (represented as ``pred $R_{AA}$'' on the $y$-axis). These values are also compared with the ``True $R_{AA}$'' on the $x$-axis, directly given by the Schr\"odinger model. Different panels represent the $R_{AA}(N_p)$ and $R_{AA}(p_T)$ of various bottomonium states.
}
  \label{fig:5020-compare-pred}
\end{figure}

\begin{figure}[htbp]
\centering
\includegraphics[width=0.48\linewidth]{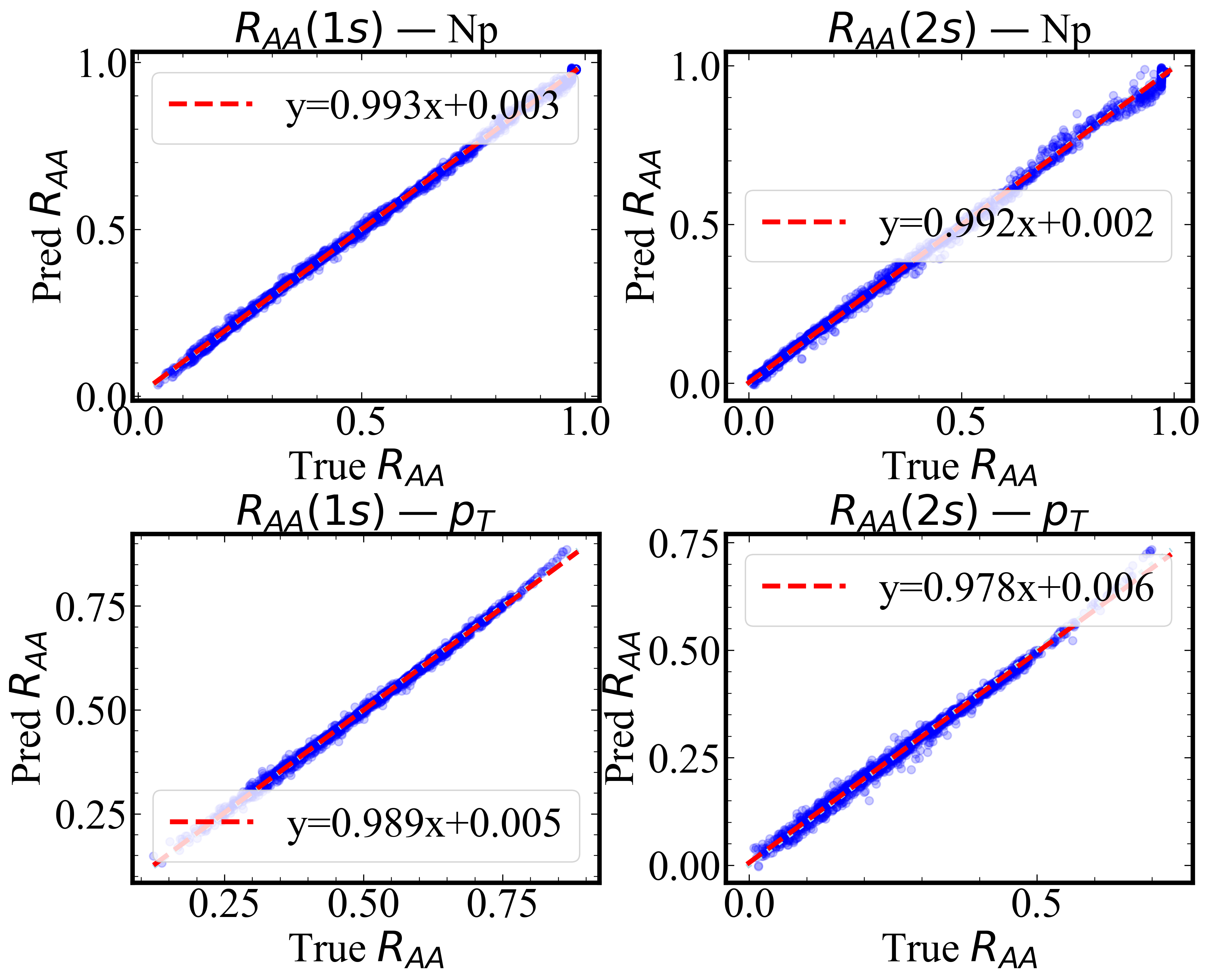}
\includegraphics[width=0.48\linewidth]{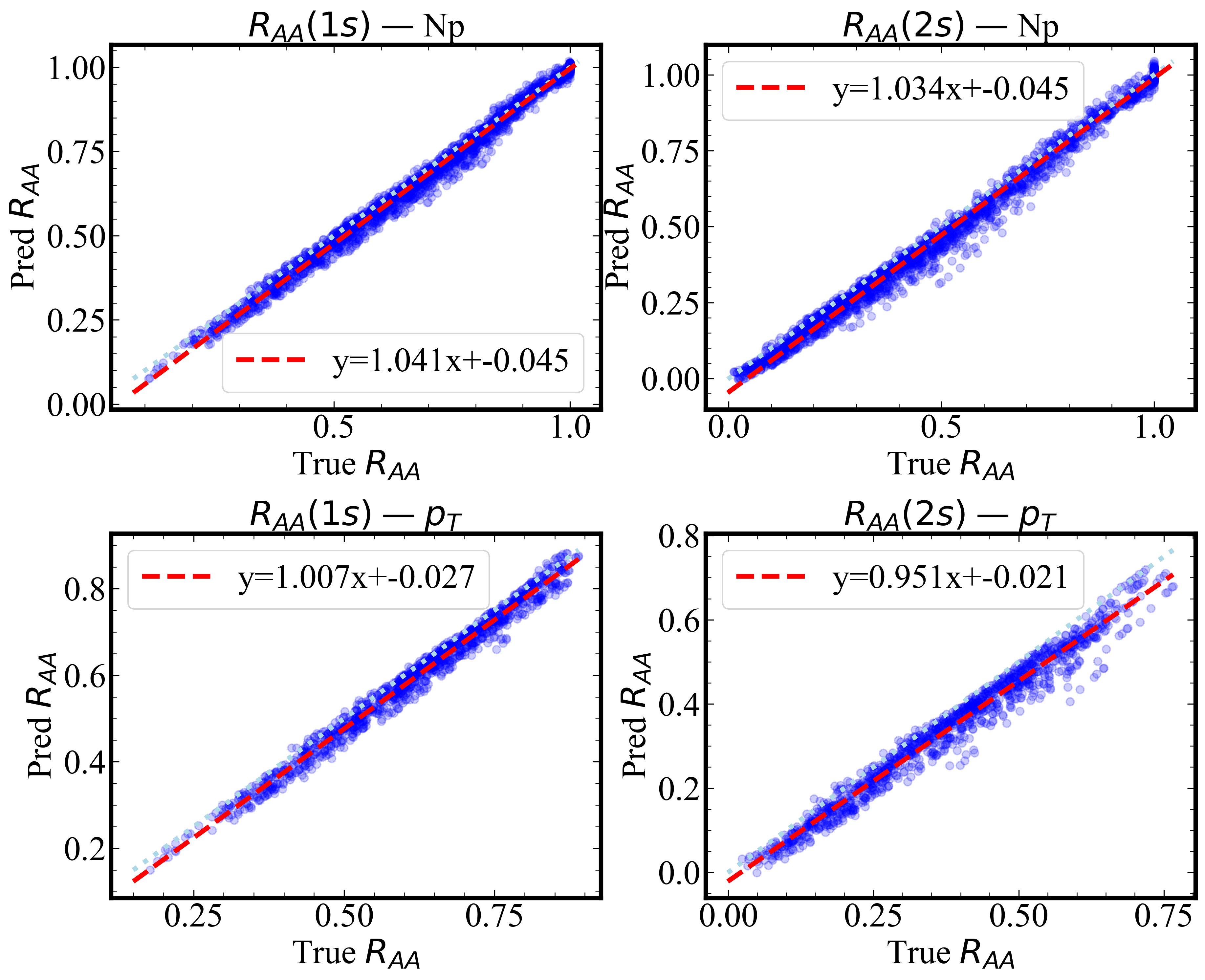}
  \caption{Left: 
  Testing performance of the CNN model at 2.76 TeV Pb-Pb collisions (left panel) and 200 GeV Au-Au collisions (right panel), respectively. Other setups are the same as Figure \ref{fig:5020-compare-pred}.}
  \label{fig:2760-compare-pred}
\end{figure}

To further evaluate the predictive performance of the trained CNNs, we utilize the independent hold-out set, consisting of 20\% of the initial 1,000 Schr\"odinger-model samples ($0.2N$) that were excluded from PCA/GP emulation and CNN training. For each parameter set $\boldsymbol{\theta}=(a_0, a_1, a_2, a_3, a_4, T_{\rm sw})$, the CNN predicts 66 distinct $R_{AA}$ values, spanning $R_{AA}(N_p)$ and $R_{AA}(p_T)$, denoted as ``predicted $R_{AA}$''. These are compared against the corresponding ground-truth results from the Schr\"odinger model, denoted as ``True $R_{AA}$''. In Figure ~\ref{fig:5020-compare-pred}, the predicted and true $R_{AA}$ values for these 200 samples (totaling $200 \times 66$ data points) are plotted as $y$ and $x$ coordinates, respectively. Linear fits for the different bottomonium states ($\Upsilon(1S), \Upsilon(2S), \Upsilon(3S)$), indicated by dashed lines, lie in proximity to the $y=x$ identity line. This alignment demonstrates that the CNN predictions are in excellent agreement with the Schr\"odinger model results. Similar performance metrics are observed for the 2.76 TeV Pb-Pb and 200 GeV Au-Au systems, as shown in Figure ~\ref{fig:2760-compare-pred}. These validations using independent hold-out data confirm that the CNN models have successfully learned the mapping from heavy quark potentials to bottomonium $R_{AA}$ observables. Consequently, these models can be reliably employed for the inverse extraction of in-medium potentials using experimental data.

\section{Extraction of in-medium heavy quark potential}
\label{sec:extract-all}

\begin{figure}[htbp]
\centering
\includegraphics[width=0.9\textwidth]{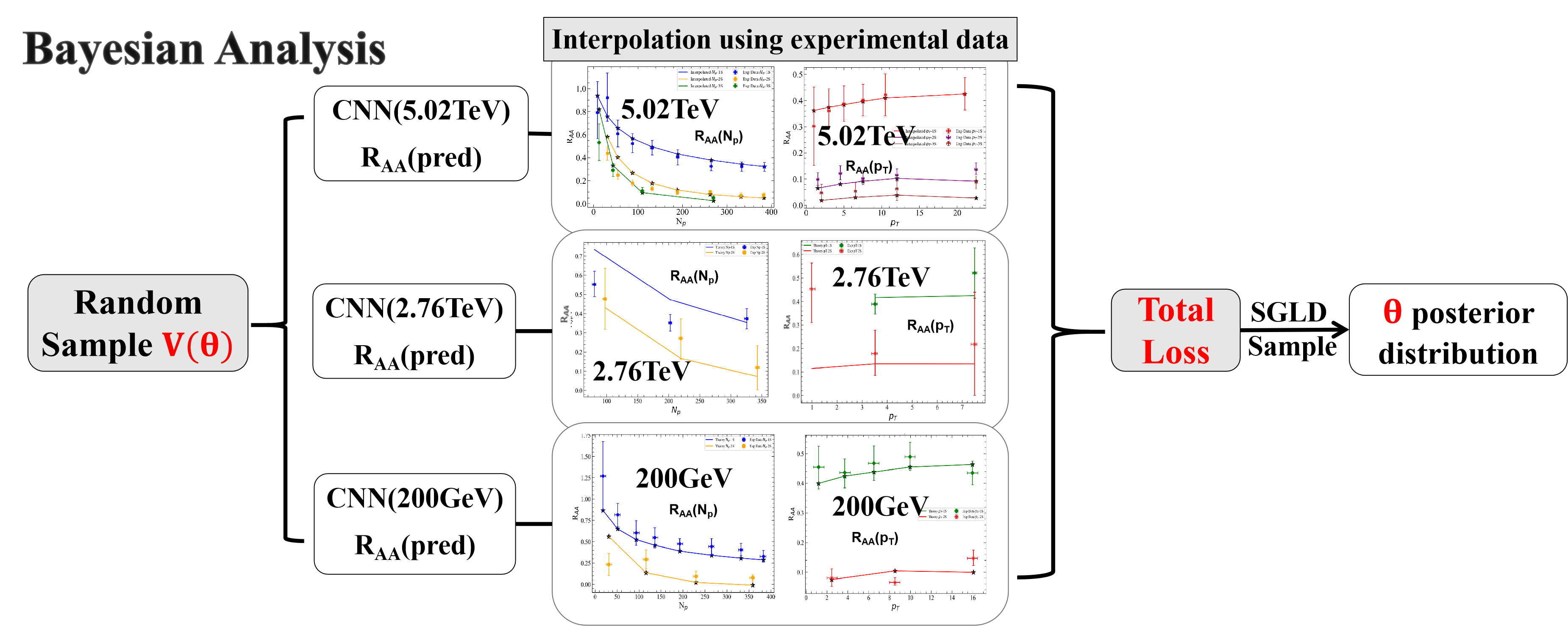}
\caption{The workflow for the SGLD-based Bayesian inversion is as follows: starting from the experimental uncertainties, we construct a diagonal covariance matrix and define the corresponding likelihood function. Posterior sampling of the potential parameters is then performed using the SGLD method. Finally, the in-medium heavy-quark potential and its uncertainty band are reconstructed from the posterior samples.}
\label{fig:Method2}
\end{figure}

In the previous sections, three independent CNNs were constructed and trained on datasets from individual collision systems to capture the mapping between bottomonium observables ($R_{AA}$) and the in-medium heavy quark potential, $V(r, T)$, characterized by the parameter vector $\boldsymbol{\theta}=(a_0, a_1, a_2, a_3, a_4, T_{\rm sw})$. In this section, we leverage these pre-trained networks to perform an inverse extraction of the parameters $\boldsymbol{\theta}$ within $V(r, T)$ by incorporating the experimental data points of bottomonium. Given that the functional form of $V(r, T)$ should remain consistent across different collision systems, we employ the framework illustrated in Figure  \ref{fig:Method2} to provide a unified description of the experimental data from all three systems simultaneously.

As illustrated in Figure  \ref{fig:Method2}, the potential parameter vector is fed in parallel into the three pre-trained CNNs to generate theoretical predictions for all channels, specifically $R_{AA}(N_p)$ and $R_{AA}(p_T)$, across the three collision systems. A total loss function is then constructed by aggregating the individual loss function from each system: $\mathcal{L}_{\rm tot}=\mathcal{L_{\rm 5.02\,TeV}}+\mathcal{L_{\rm 2.76\,TeV}}+\mathcal{L_{\rm 200\,GeV}}$, where the loss function for each specific system is defined in Eq.(\ref{eq:weighted-loss}). The extraction of $V(r,T)$ is treated as an inverse problem, where $\mathcal{L}_{\rm tot}$ is minimized by iteratively updating the parameter vector $\boldsymbol{\theta}$. As the total loss function $\mathcal{L}_{\rm tot}$ converges to a stable minimum, we identify the optimal potential parameters that best describe the experimental data across all three collision systems simultaneously. However, even at the global minimum of the loss function, the optimal parameter set may not be unique; multiple parameter combinations can yield comparable levels of agreement with the data. This inherent non-uniqueness, often referred to as parameter degeneracy, implies that the potential is better characterized by probability distributions rather than unique point estimates. Furthermore, the total uncertainty in $\boldsymbol{\theta}$ is compounded by experimental error bars and the architectural constraints of the CNN models. Consequently, in the following section, we report both the maximum a posteriori (MAP) estimates and the corresponding uncertainty intervals for each parameter.

\subsection{Stability of extracting heavy quark potentials}
\label{sec:stability}

\begin{figure}[htbp]
  \centering
\includegraphics[width=0.9\textwidth]{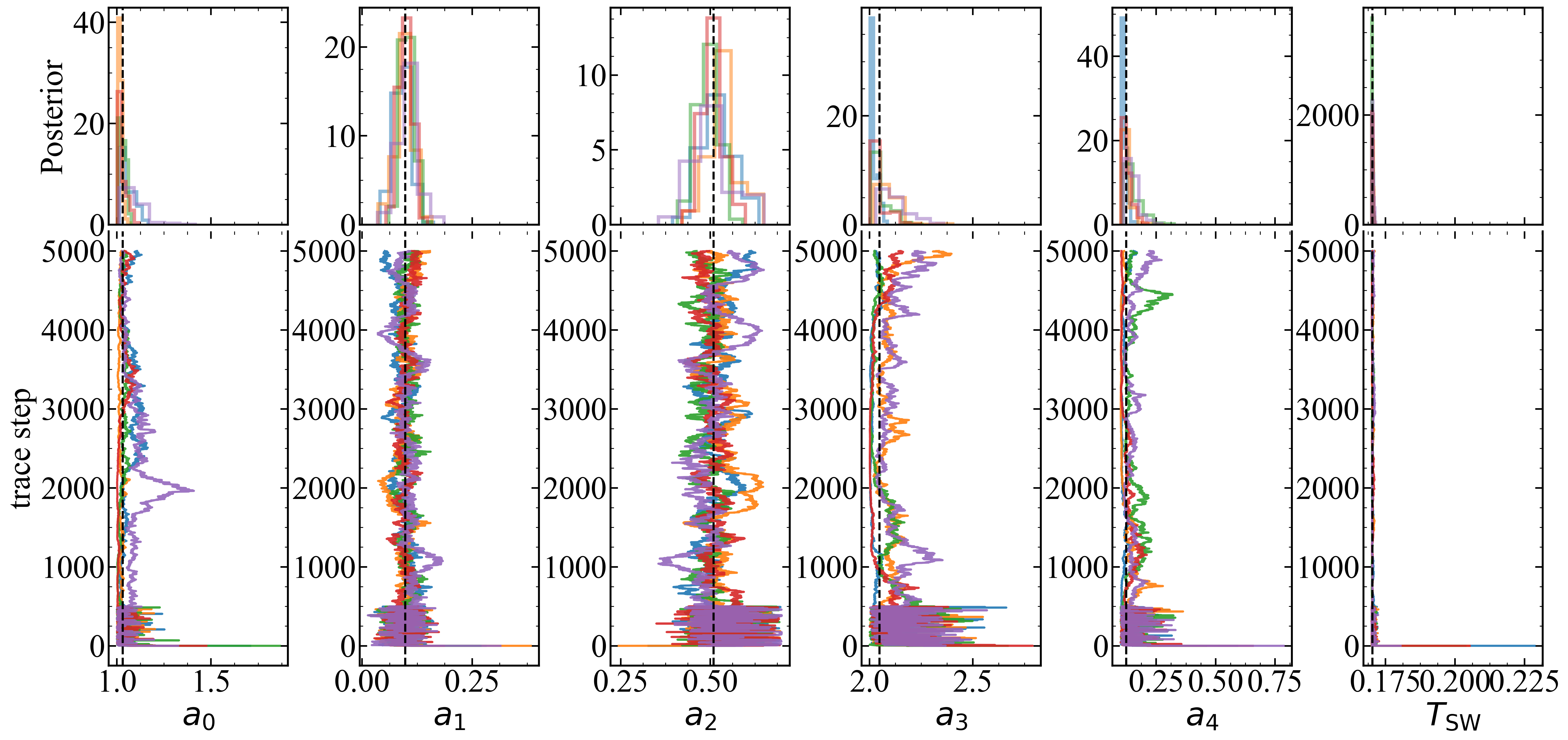}
  \caption{Marginal posterior distributions (upper panels) and corresponding MCMC trajectories (lower panels) of the heavy quark potential parameters $\boldsymbol{\theta}$. The upper panels display the posterior distributions extracted from bottomonium $R_{AA}$ experimental data across three collision systems (5.02 TeV Pb-Pb, 2.76 TeV Pb-Pb, and 200 GeV Au-Au), while the lower panels show the corresponding sampling trajectories. Five distinct colors represent results from independent Markov chains.
}
  \label{lab-fig:three-5trace}
\end{figure}

To assess the stability and robustness of the CNN models, we employ 5 independent parallel chains to track the iteration process of the parameter vector $\boldsymbol{\theta}$ during the potential extraction process. Each chain is initialized independently in the latent space $\boldsymbol{z}$, where $\boldsymbol{\theta} = \boldsymbol{\theta}(\boldsymbol{z})$. By comparing the resulting posterior distributions from these 5 chains, we can evaluate the convergence of the sampling process; consistent distributions across all chains demonstrate the reliability of the CNN-based framework in extracting the in-medium heavy quark potential $V(r, T)$. Posterior sampling of $\boldsymbol{\theta}$ is conducted using the Stochastic Gradient Langevin Dynamics algorithm. As illustrated in Figure \ref{lab-fig:three-5trace}, each chain undergoes $5 \times 10^6$ iterations to minimize the total loss $\mathcal{L}_{\rm tot}$. To ensure the quality of the samples, we discard the first $5 \times 10^4$ iterations as a burn-in period and apply a thinning factor of 20 to mitigate autocorrelation between successive draws. This procedure yields approximately 1.24 million retained samples $\{\boldsymbol{\theta}\}$, which are utilized to approximate the joint posterior distribution of the potential parameters. The parameter set that minimizes the total loss function $\mathcal{L}_{\rm tot}$ is identified as the Maximum A Posteriori (MAP) estimate, denoted as $\widehat{\boldsymbol{\theta}}_{\rm MAP}$. This estimate represents the potential configuration that provides the best simultaneous explanation of the bottomonium $R_{AA}$ experimental data across all three collision systems. As shown in the upper small panels of Figure \ref{lab-fig:three-5trace}, the marginal posterior distributions (5 lines) of each potential parameter are independent of their initializations, indicating that the CNN models and the integrated framework successfully converge to a consistent representation of the heavy quark potential.

\subsection{In-medium heavy quark potentials}
\label{sec:in-medium-HQ-V}

\begin{figure}[htbp]
  \centering
\includegraphics[width=0.9\textwidth]{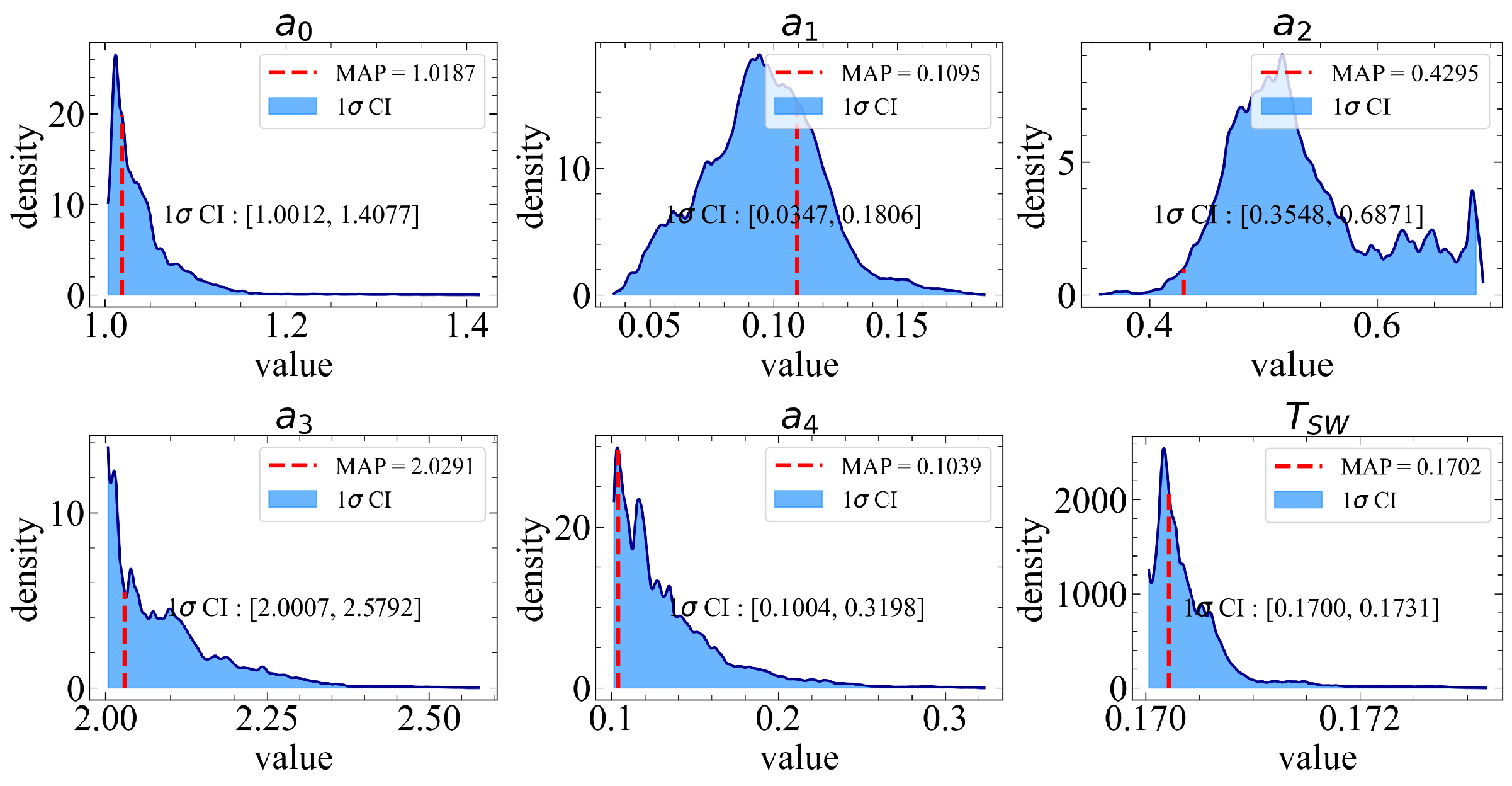}
  \caption{Joint Highest Posterior Density (HPD) projections for the six potential parameters. Each panel displays the marginal posterior distribution (color band), representing the $1\sigma$ uncertainty for a specific parameter, with the red dashed line indicating the Maximum A Posteriori (MAP) value.  }
  \label{lab-fig:three-parap-HPD2}
\end{figure}

From over 1 million retained samples generated across the 5 independent chains, the Maximum A Posteriori (MAP) estimate of the potential parameters, $\widehat{\boldsymbol{\theta}}_{\rm MAP}$, is determined. The in-medium heavy quark potential derived from $\widehat{\boldsymbol{\theta}}_{\rm MAP}$ represents the optimal configuration for interpreting the bottomonium $R_{AA}$ experimental data across all considered systems. This optimal potential serves as the primary baseline for our analysis, as illustrated in Figure  \ref{lab-fig:three-parap-HPD2}, where the values corresponding to $\widehat{\boldsymbol{\theta}}_{\rm MAP}$ are indicated by dashed lines.

The uncertainties of the potential parameters are quantified based on the distributions shown in Figure  \ref{lab-fig:three-parap-HPD2}. To determine these uncertainty intervals, we rank over 1 million retained SGLD samples according to their negative log-likelihood (NLL), $\mathcal{L}(\boldsymbol{\theta})$, in ascending order. We then define the joint Highest Posterior Density (HPD) set, $\mathcal{S}_\gamma= \bigl\{\boldsymbol{\theta}\bigr\}$ with top $\gamma=68.27\%$ of the retained $\mathcal{L}_{\rm tot}(\boldsymbol{\theta})$-ranked samples. The parameter sets within $\mathcal{S}_\gamma$ maintain a high level of agreement with the bottomonium experimental data across all three collision systems simultaneously. These selected samples are utilized to characterize the $1\sigma$ uncertainty intervals for the in-medium heavy quark potential $V(r, T)$. Both the optimal MAP values and the corresponding uncertainties for each parameter are summarized in Table \ref{tab:parameter-intervalsp2}.

\newcolumntype{C}[1]{>{\centering\arraybackslash}m{#1}} 
\begin{table}[htp]
\centering
\caption{ The uncertainties and MAP values of the potential parameters extracted with NLL-based joint HPD. The top $68.27\%$ samples $\boldsymbol{\theta}$ from the SGLD process are retained to approximate the joint HPD set $\mathcal{S}_{\gamma}$. The real and imaginary parts of the in-medium heavy quark potential are also shown for clarity. }
\label{tab:parameter-intervalsp2}
\renewcommand{\arraystretch}{1.2} % 
\setlength{\tabcolsep}{6pt} % 

\begin{tabular}{|C{2.5cm}|C{3.5cm}|C{3cm}|}  % 使用自定义列类型
\hline
\textbf{\normalsize Parameters} & \textbf{\normalsize $1\sigma$ Uncertainties} & \textbf{\normalsize MAP Value} \\
\hline
\footnotesize $a_0$ & \footnotesize [1.001,\,1.407] & \footnotesize 1.018 \\
\footnotesize $a_1$ & \footnotesize [0.034,\,0.180] & \footnotesize 0.109 \\
\footnotesize $a_2$ & \footnotesize [0.354,\,0.687] & \footnotesize 0.429 \\
\footnotesize $a_3$ & \footnotesize [2.000,\,2.579] & \footnotesize 2.029 \\
\footnotesize $a_4$ & \footnotesize [0.100,\,0.319] & \footnotesize 0.103 \\
\footnotesize $T_{\text{sw}}$ & \footnotesize [0.170,\,0.173] & \footnotesize 0.170 \\
\hline
\multicolumn{3}{|c|}{\textbf{\normalsize Potential parametrization:}} \\
\multicolumn{3}{|C{9.6cm}|}{  % 改为 C 类型，保持居中
  \vspace{1pt} % 减少垂直间距
  \centering  % 确保多行公式整体居中
  \footnotesize % 公式部分使用较小字体
  $\displaystyle V_I = -\,i\,T^{a_{0}} \bigl(a_1 r + a_2 r^{a_3}\bigr)$ \\[2pt] % 减小行间距
  $\displaystyle V_R = -\alpha\!\left(\mu_D + \frac{e^{-\mu_D r}}{r}\right) + \frac{2\sigma}{\mu_D} - \frac{\sigma e^{-\mu_D r}\,(2 + \mu_D r)}{\mu_D}$ \\[2pt] % 减小行间距
  $\displaystyle \text{with Debye mass } \mu_D = a_4\,T \sqrt{\frac{4\pi N_c}{3}\!\left( 1 + \frac{N_f}{6} \right)\alpha}$ 
  \vspace{1pt} % 减少垂直间距
} \\
\hline
\end{tabular}
\end{table}

In Table~\ref{tab:parameter-intervalsp2}, the parameter $a_0$ and the set $(a_1, a_2, a_3)$ characterize the temperature and spatial dependencies, respectively, of the imaginary part of the heavy quark potential, $V_I(T, r)$. The parameter $a_4$ governs the temperature dependence of the Debye mass within the real part of the potential, $V_R(T, r)$. Additionally, $T_{\rm sw}$ denotes the switching temperature at which the in-medium parametrization transitions to the vacuum Cornell potential $V_c(r)$, as previously described. Notably, the Maximum A Posteriori (MAP) estimate for $T_{\rm sw}$ is in excellent agreement with the expected pseudo-critical temperature $T_c$. This consistency provides a strong validation that the proposed parametrization in Table~\ref{tab:parameter-intervalsp2} is sufficiently robust to accurately describe the in-medium heavy quark potential across the entire range of the Schr\"odinger equation's evolution.

\begin{figure}[htbp]
  \centering
\includegraphics[width=0.9\textwidth]{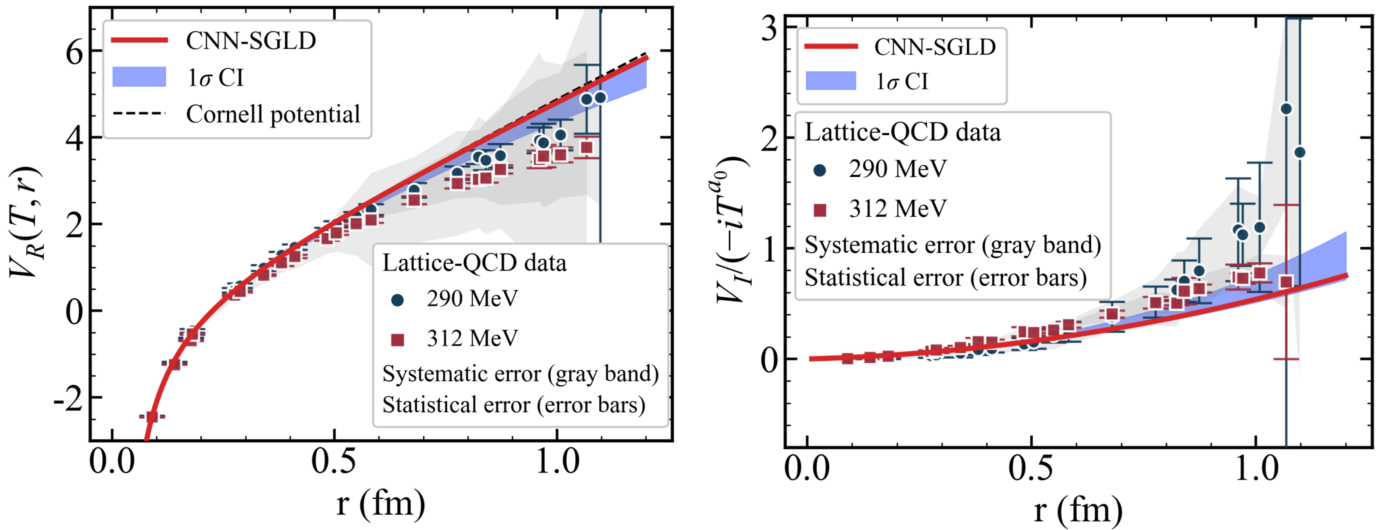}
  \caption{Optimal parametrization (thick solid lines) and $1\sigma$ uncertainty intervals (color bands) of the in-medium heavy quark potential, extracted via the CNN-SGLD framework. The extraction is based on experimental data for bottomonium $R_{AA}(1S, 2S, 3S)$ across three collision systems: 5.02 TeV Pb-Pb, 2.76 TeV Pb-Pb, and 200 GeV Au-Au. The vacuum Cornell potential is indicated by the thin dotted line for comparison. The left and right panels display the real part $V_R$ and the scaled imaginary part $V_I/(-iT^{a_0})$ of the potential, respectively. Lattice QCD results are from Ref.~\cite{Burnier_2017}.}
  \label{fig:three-potential2}
\end{figure}

In the real component of the in-medium heavy quark potential, the color screening effect is expected to increase with temperature and is characterized by the Debye mass, $\mu_D$. We adopt the functional form derived from Hard Thermal Loop (HTL) calculations, incorporating an adjustable scaling factor $a_4$ within $\mu_D$. In the limit $a_4 \to 0$, color screening becomes negligible, and $V_R$ approaches the vacuum Cornell potential. As shown in Table~\ref{tab:parameter-intervalsp2}, the optimal $a_4$ value extracted by the CNN models is approximately 0.1. This relatively small value suggests a weak color screening effect, which is qualitatively consistent with recent lattice QCD results~\cite{Bazavov:2023dci} and Bayesian analyses~\cite{Burnier:2013nla,Bzdak:2017ltv,Schenke:2020uqq}. For the imaginary potential $V_I(T,r)$, the optimal value of $a_0$ is near 1.0, suggesting that $V_I$ depends linearly on temperature. We parametrize the spatial dependence of the imaginary part using a polynomial expansion in $r$. To minimize the number of free parameters, we employ a two-term form: $V_I(r)/(-iT^{a_0}) = a_1 r + a_2 r^{a_3}$. Here, the second term ($a_2 r^{a_3}$) captures higher-order contributions (e.g., $r^2, r^3$); consequently, the sampling range for the exponent $a_3$ is constrained to $a_3 \ge 2.0$. The optimal value for $a_3$ is close to 2.0, indicating that terms up to $r^2$ are sufficient for the parametrization. While a purely linear $r$-dependence would require the coefficient $a_2$ to vanish, our optimal estimate of $a_2 = 0.4295$ (within the sampling range $a_2\in[0.2, 0.7]$) confirms that higher-order terms are necessary to accurately describe the bottomonium $R_{AA}$s. Using `\textit{Maximum A Posteriori}'  estimation, the optimal real and imaginary potentials, along with their $1\sigma$ uncertainties, are illustrated in Figure ~\ref{fig:three-potential2}. Take a typical temperature $T=300$ MeV available at LHC energies as an example, the thick solid lines represent the potential that simultaneously provides the best description of bottomonium data across three collision systems (5.02 TeV Pb-Pb, 2.76 TeV Pb-Pb, and 200 GeV Au-Au). Notably, the MAP result for the real potential remains remarkably close to the vacuum Cornell potential (indicated by the thin dotted line in the left panel of Figure \ref{fig:three-potential2}), which also agrees with recent deep learning heavy quark potential reconstruction from LQCD results on in-medium heavy quarkonium spectroscopy\cite{Shi:2021qri}. The shaded bands represent the $1\sigma$ uncertainties corresponding to the parameter samples $\mathcal{S}_\gamma$. The color band of the imaginary potential covers most of the lattice QCD data points~\cite{Burnier_2017}.

\subsection{Model validation with mock data}

\begin{figure}[htp]%[H]
  \centering
\includegraphics[width=0.8\linewidth]{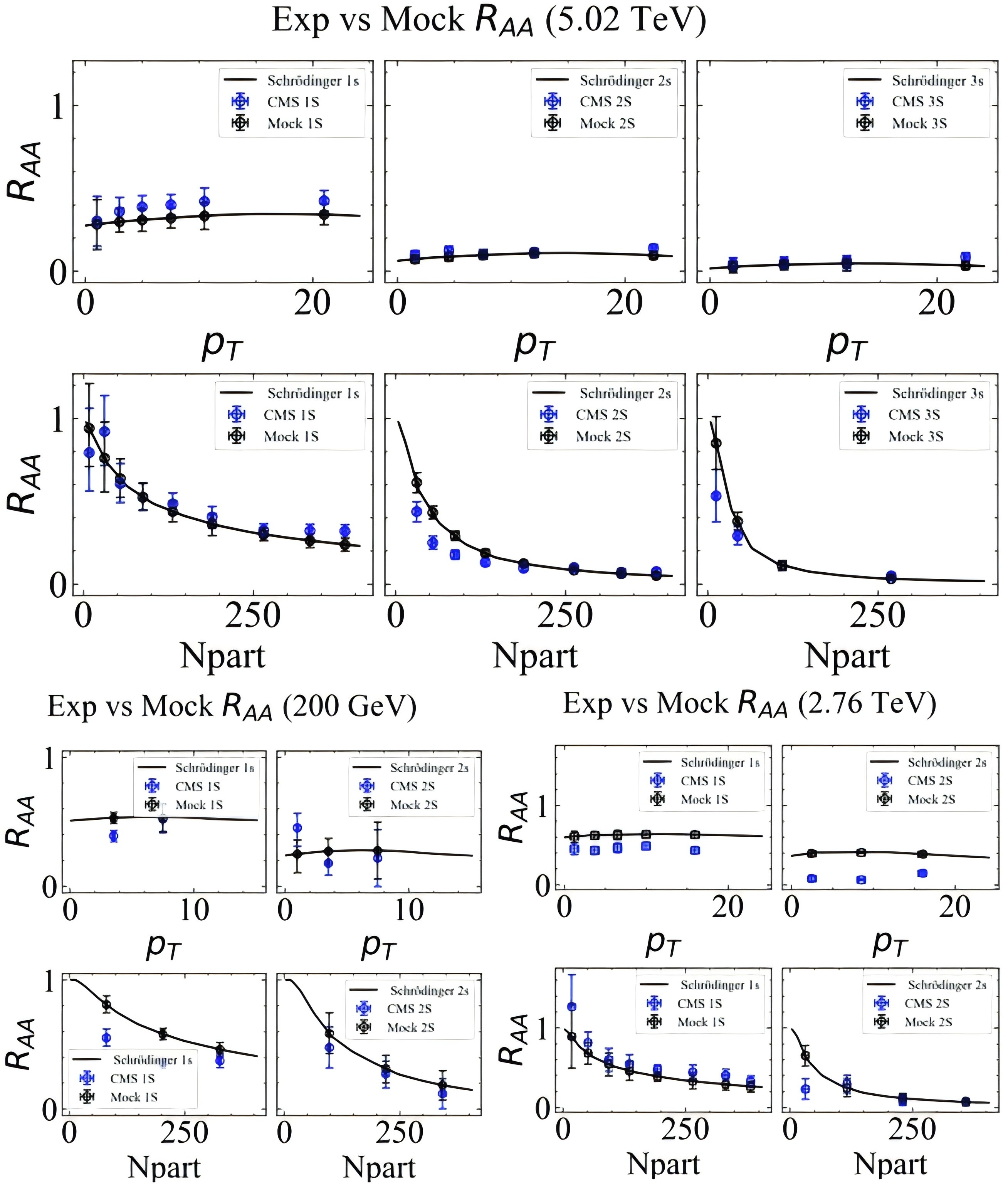}
\caption{Mock-data validation across three collision energies. Black solid curves represent the results calculated using the Schr\"odinger model with the mock potential parameters $\boldsymbol{\theta}^{\rm mock}$ defined in Eq.~(\ref{lab-mockV}). Discrete synthetic data (black dots) are sampled from these theoretical curves to match the binning and structure of the experimental data (blue open circles).
}
  \label{fig:mock-one-group2}
\end{figure}

To validate the reliability of our extraction procedure, we performed a closure test using synthetic data. First, we randomly sampled a parameter set, $\boldsymbol{\theta}^{\rm mock}$, and incorporated the resulting parametrized potential, $V^{\rm mock}(T,r;\boldsymbol{\theta}^{\rm mock})$, into our Schr\"odinger model to calculate the corresponding nuclear modification factors. The results of these calculations are shown as solid lines in Figure ~\ref{fig:mock-one-group2}, alongside the actual experimental data points (blue markers). To ensure the theoretical output is structurally consistent with experimental observations, we sampled discrete points from the calculated theoretical curves. These synthetic data points, indicated by black dots in Figure ~\ref{fig:mock-one-group2}, constitute the mock $R_{AA}$ dataset for the $\Upsilon(1S, 2S, 3S)$ states across each collision system. Since the CNN-SGLD framework requires experimental uncertainties as input, each mock data point was assigned the same statistical and systematic errors as its corresponding experimental counterpart, as represented by the error bars on the black points. Consequently, we established a mapping of $V^{\rm mock}(T,r;\boldsymbol{\theta}^{\rm mock}) \mapsto R_{AA}^{\rm mock}(1S,2S,3S)$ based on the Schr\"odinger model. These mock observations, $R_{AA}^{\rm mock}$, were then fed into the pre-trained CNNs and SGLD framework to inversely extract the potential parameters $\boldsymbol{\theta}_\star$. Finally, the extracted values $\boldsymbol{\theta}_\star$ were compared against the ``ground truth'' values $\boldsymbol{\theta}^{\rm mock}$ to verify the accuracy of the extraction.

A representative sample of the heavy quark potential parameters is randomly generated as: 
\begin{equation} \label{lab-mockV} \boldsymbol{\theta}^{\rm mock} = (a_0, a_1, a_2, a_3, a_4, T_{\rm sw}) = (1.6 , 0.2 , 0.3 , 2.7 , 0.3 , 0.18), 
\end{equation} 
where the chosen values lie within the prior parameter ranges established in the preceding analysis.

\begin{figure}[htbp]
  \centering
\includegraphics[width=0.9\linewidth]{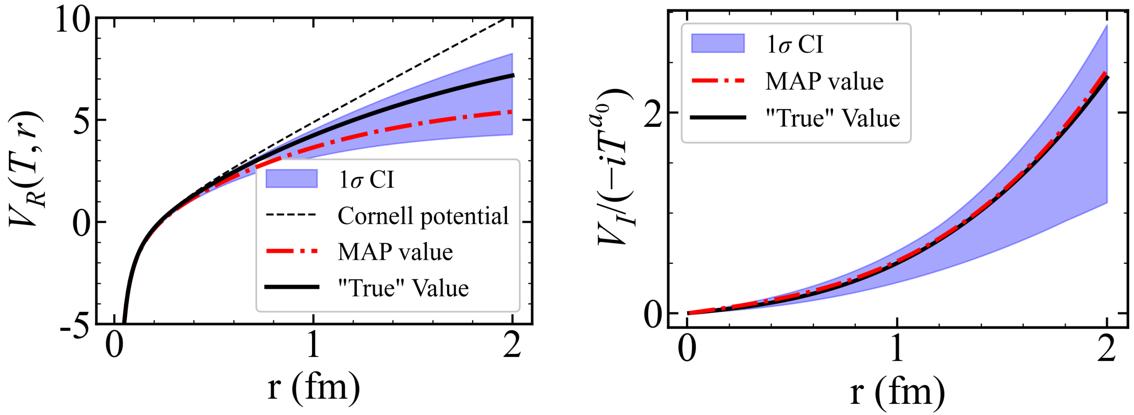}
\caption{The real (left) and imaginary (right) components of the reconstructed heavy quark potential $V_\star(T,r;\boldsymbol{\theta}_\star)$, extracted via the trained CNN-SGLD framework, are shown as red dotted-dashed lines. The shaded regions represent the $1\sigma$ credible intervals. The ground-truth potential, $V^{\rm mock}(T,r; \boldsymbol{\theta}^{\rm mock})$, is indicated by the black solid lines for comparison.}
  \label{lab-fig:mock-V-com}
\end{figure}

By inputting the synthetic data $R_{AA}^{\rm mock}$ into the deep neural networks, we extract the corresponding potential parameters $\boldsymbol{\theta}_\star$ and the in-medium heavy quark potential $V_\star(T, r; \boldsymbol{\theta})$ via the SGLD process. In Figure ~\ref{lab-fig:mock-V-com}, the  MAP (Maximum A Posteriori) curves of the extracted potential, $V_\star(T, r; \boldsymbol{\theta}_\star)$, is represented by red dotted-dashed lines, with the shaded bands denoting their $1\sigma$ credible intervals. For comparison, the ``ground truth'' potential $V^{\rm mock}(T, r; \boldsymbol{\theta}^{\rm mock})$ is plotted as black solid lines. Regarding the imaginary components, the black and red curves exhibit near-perfect overlap. These results demonstrate that the trained CNN-SGLD framework can accurately reconstruct the underlying mapping between the in-medium potential $V(T, r)$ and bottomonium observables.

\subsection{Recalculate Bottomonium $R_{AA}$ with the extracted
potentials}
\begin{figure}[htp]%[H]
  \centering
\includegraphics[width=0.8\linewidth]{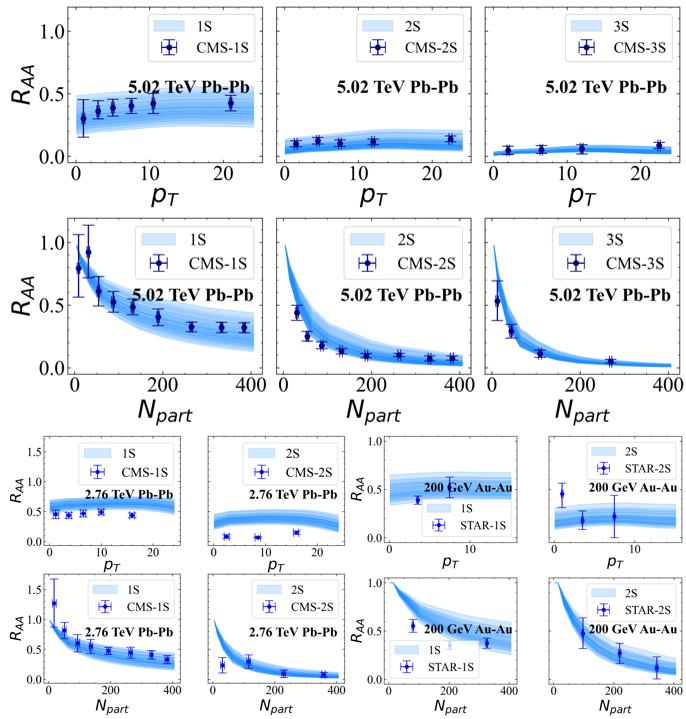}
  \caption{To quantify the uncertainty in our results, 200 parameter sets were sampled from the $1\sigma$ credible intervals (CI) presented in Table~\ref{tab:parameter-intervalsp2}. These potential samples were then incorporated into the Schr\"odinger equation framework to calculate the bottomonium $R_{AA}$ for the three collision systems under consideration. }
  \label{fig:sample-new-para}
\end{figure}

Finally, the extracted in-medium heavy quark potentials from Section~\ref{sec:in-medium-HQ-V}  are incorporated back into the Schr\"odinger equation framework to calculate the corresponding bottomonium nuclear modification factors. These results are compared with the experimental measurements in Figure ~\ref{fig:sample-new-para}. The reconstructed potentials from Section~\ref{sec:in-medium-HQ-V} demonstrate a strong level of agreement, successfully reproducing the majority of the bottomonium $R_{AA}$ data points across the various collision systems.

\section{Summary}
In this study, we employ a deep learning framework to extract the in-medium heavy quark potential from bottomonium observables across multiple heavy ion collision systems, specifically 5.02 TeV and 2.76 TeV Pb-Pb, and 200 GeV Au-Au collision systems. We develop a Convolutional Neural Network architecture to establish a functional mapping from the parameterized in-medium potential (input) to the nuclear modification factors ($R_{AA}$) of various bottomonium eigenstates (output). The training datasets are generated by solving the time-dependent Schr\"odinger equation, where the real and imaginary components of the potential are systematically varied and integrated with a hydrodynamically evolving bulk medium. To obtain more events in the datasets from the Schr\"odinger model, Principal Component Analysis and a Gaussian Process (GP) emulator are utilized. Following the independent training of CNNs for each collision system, we implement the Stochastic Gradient Langevin Dynamics (SGLD) to inversely extract the heavy quark potential parameters by incorporating experimental measurements of bottomonium. The optimal parameters are determined by minimizing a joint loss function across all three systems, with $1\sigma$ uncertainties quantified by accounting for both experimental errors and the intrinsic variance of the CNN model involved. The reliability of this CNN-SGLD framework is validated through closure tests using mock data, which demonstrates the high robustness and effectiveness of the extraction pipeline. Our findings under the current heavy ion collision measurements indicate that the real part of the in-medium potential remains remarkably close to the vacuum Cornell potential, suggesting that color screening is negligible within the temperature ranges accessible in these collision systems. In contrast, the imaginary component emerges significantly and dominates the bottomonium suppression from RHIC to LHC energies.

While a unified potential $V(T, r)$ is derived from a simultaneous fit to the data from all three collision energies, we also explored an alternative approach in the Appendix: extracting $V(T, r)$ independently for each system by minimizing their corresponding loss function. A comparison revealed that the $1\sigma$ credible intervals for the independently extracted parameters exhibited noticeable discrepancies across systems. The lack of convergence across the three collision systems underscores the challenge of deriving a unified functional form from experimental data points of a single system alone, confirming the necessity of using multi-energy bottomonium data to robustly constrain the in-medium potential.

\clearpage
\begin{appendices}
\renewcommand{\thesection}{\Roman{section}}  
\renewcommand{\thesubsection}{\Alph{subsection}.}  

\section{Appendix}
\subsection{Extracting heavy quark potential in each collision system}

\begin{figure}[htp]
  \centering
\includegraphics[width=0.9\textwidth]{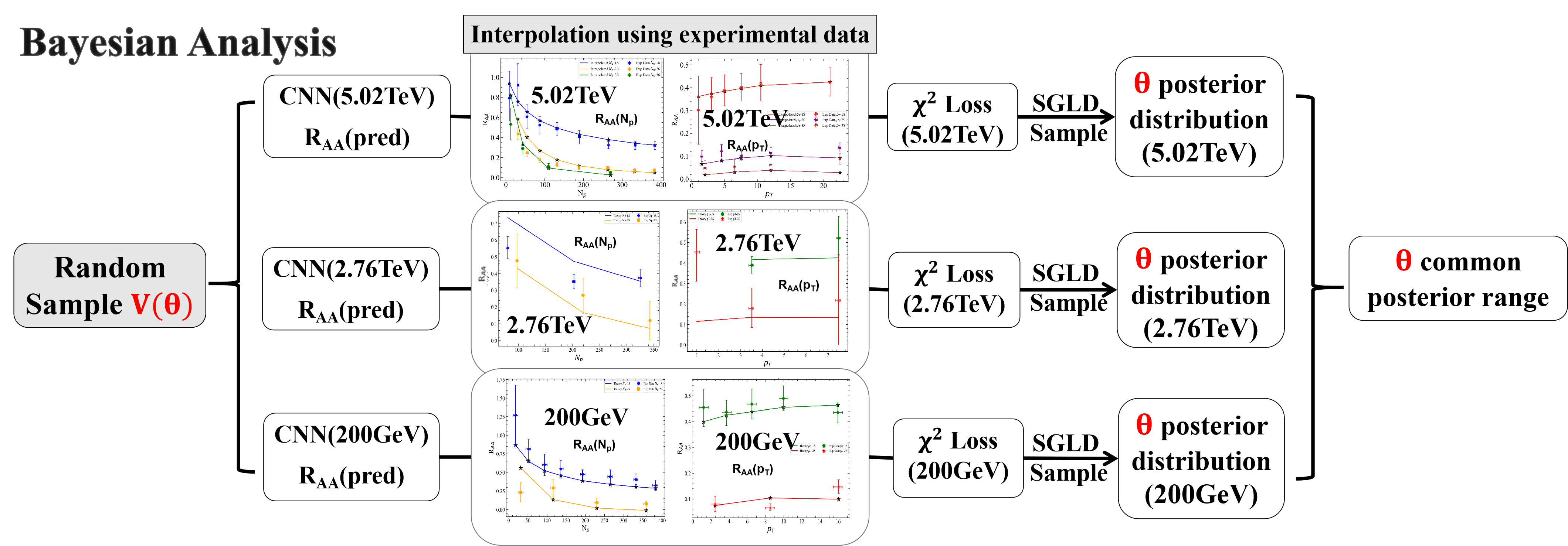}
  \caption{Schematic representation of the inverse extraction of the in-medium heavy quark potential using CNNs and the SGLD process. The extraction is performed independently for each of the three collision systems: 5.02 TeV Pb-Pb, 2.76 TeV Pb-Pb, and 200 GeV Au-Au. The procedure highlights the overlap (or discrepancy) among the three reconstructed potential parameter vectors.}
  \label{lab-workflow-met1}
\end{figure}

\begin{figure}[htbp]
  \centering
\includegraphics[width=0.8\linewidth]{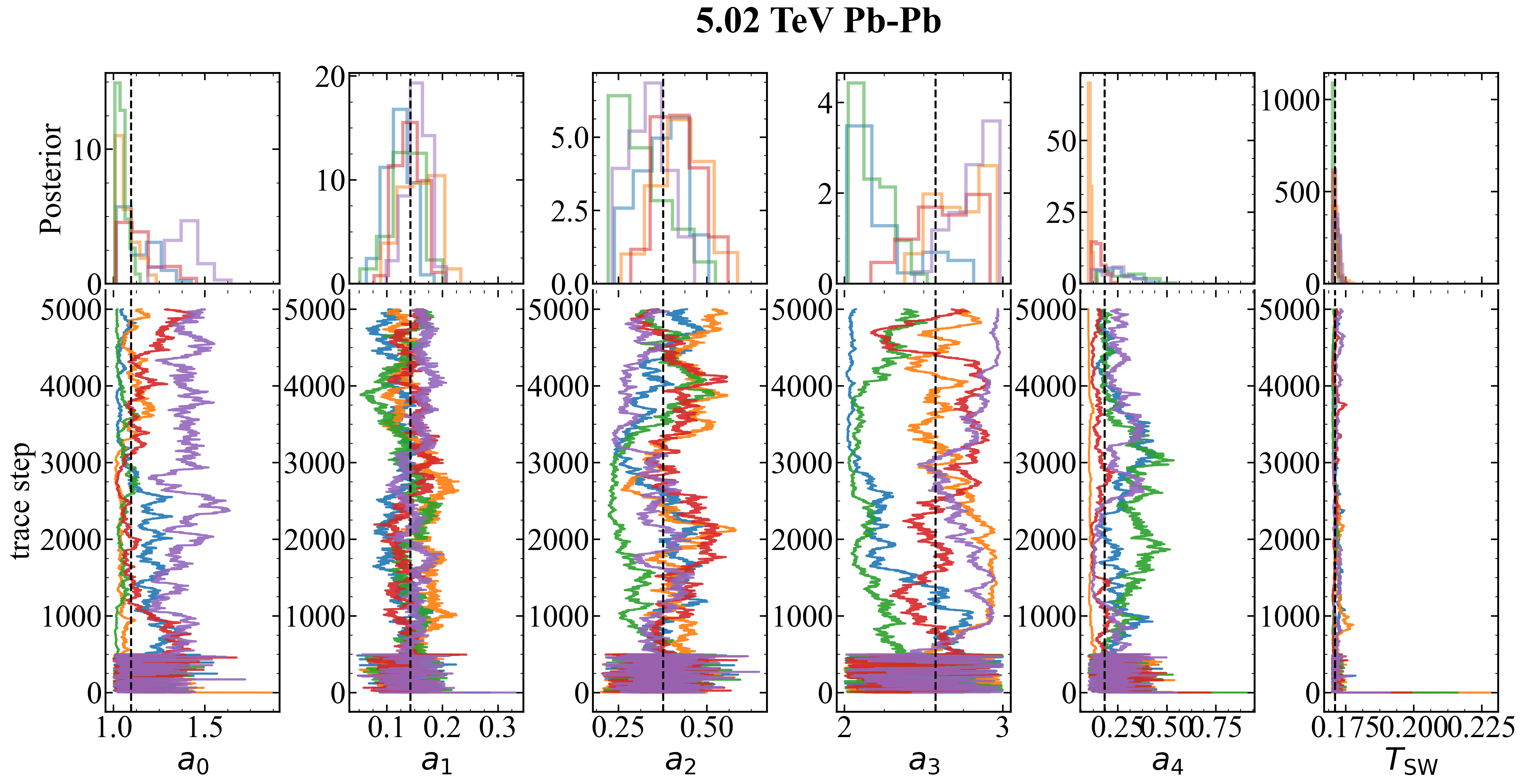}
\includegraphics[width=0.8\linewidth]{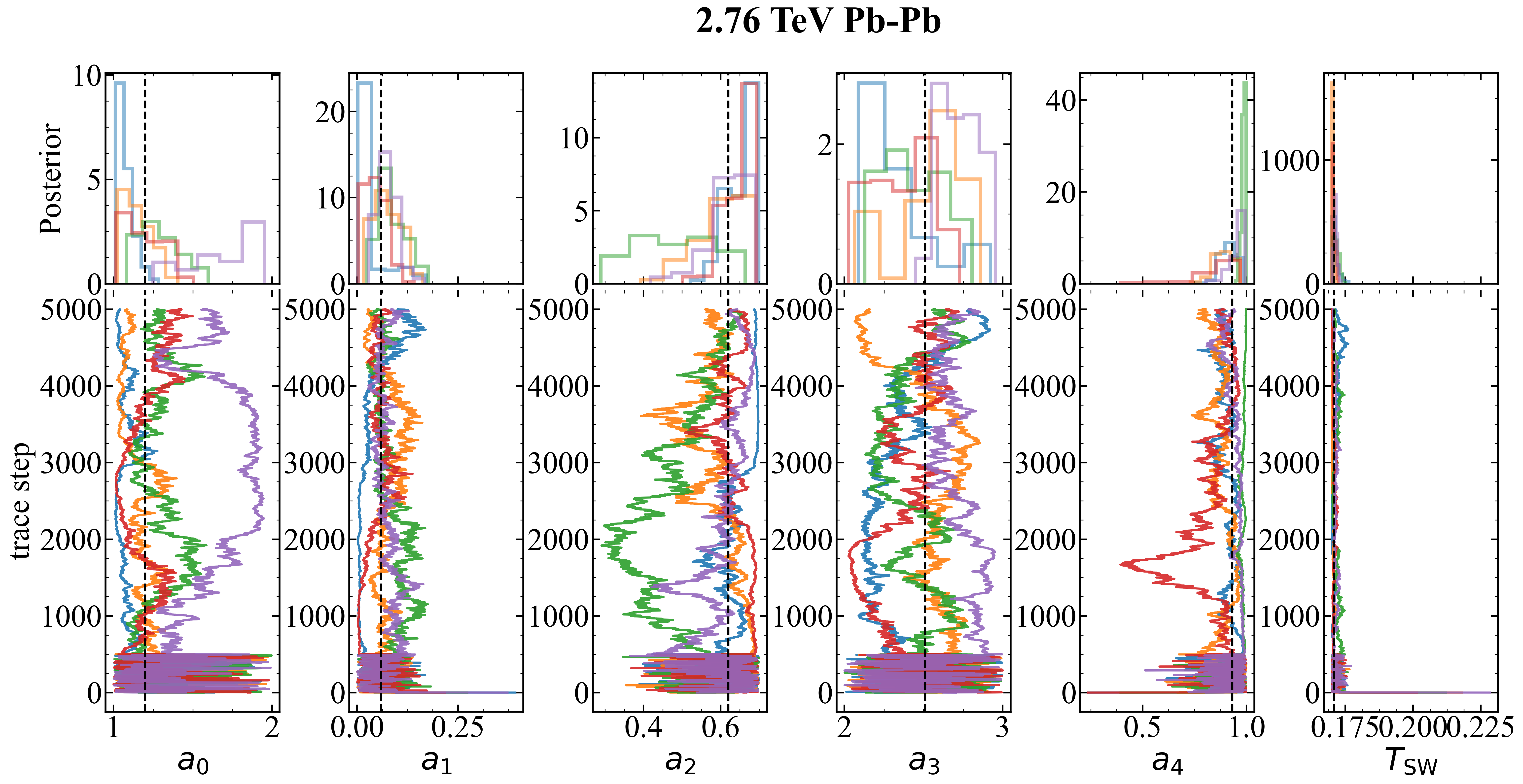}
\includegraphics[width=0.8\linewidth]{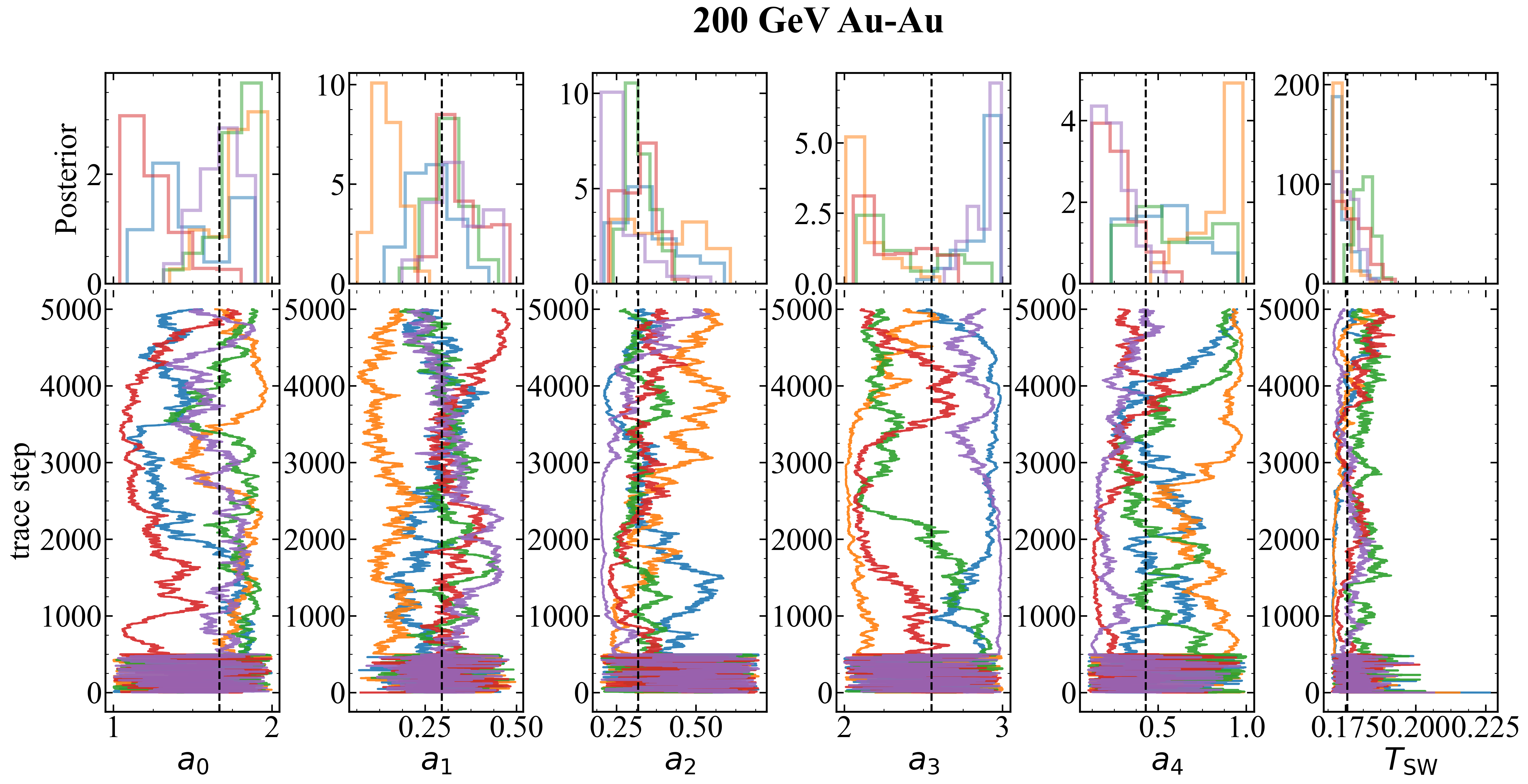}
  \caption{ Marginal posterior distributions (upper panel) and simulated trajectories (lower panel) of the potential parameters $\boldsymbol{\theta}$, reconstructed independently from the experimental $R_{AA}$ data of various bottomonium eigenstates across three collision systems: $5.02~\text{TeV}$ Pb-Pb, $2.76~\text{TeV}$ Pb-Pb, and $200~\text{GeV}$ Au-Au. In the lower panel, the different colored lines represent the evolution of parameters within each Markov chain during the SGLD process. The definition of the marginal posterior distributions remains consistent with the description in previous sections.
}
  \label{fig:3-5-trace-app}
\end{figure}

For each collision system, a specific Convolutional Neural Network (CNN) is constructed and trained independently using datasets generated by the Schr\"odinger equation tailored to that system. 
In this appendix, we explore an alternative approach. While the CNNs are still trained on system-specific datasets, we perform independent extractions of the heavy quark potential using the experimental data points from each collision system separately. In this framework shown as Figure \ref{lab-workflow-met1}, each potential is determined by minimizing the individual loss function corresponding to its specific collision system. Therefore, three distinct in-medium heavy quark potentials are extracted independently through the SGLD-based training of three separate CNN architectures. An intersection among the three independently extracted potentials would be considered an optimal configuration, potentially providing a simultaneous description of bottomonium $R_{AA}$ across all three collision systems. Similar to Section \ref{sec:stability},
Figure \ref{fig:3-5-trace-app} displays the marginal posterior distributions and simulated trajectories of the potential parameters $\boldsymbol{\theta}$, reconstructed independently from the bottomonium nuclear modification factor ($R_{AA}$) data at each of three collision energies. To construct the $1\sigma$ credibility interval (CI) for $\boldsymbol{\theta}$, we sort the retained samples by their loss function values in ascending order and select the 68.27\% of samples with the smallest values. As shown in the top panel of Figure \ref{fig:3-5-trace-app}, there are evident discrepancies between the marginal posterior distributions of ($a_2, a_3$), which govern the distance dependence of the imaginary part of the heavy quark potential. In the lower panel, the different colored lines represent the results from distinct Markov chains for each collision system.

\begin{figure}[htp]%[H]
  \centering
  \includegraphics[width=0.8\linewidth]{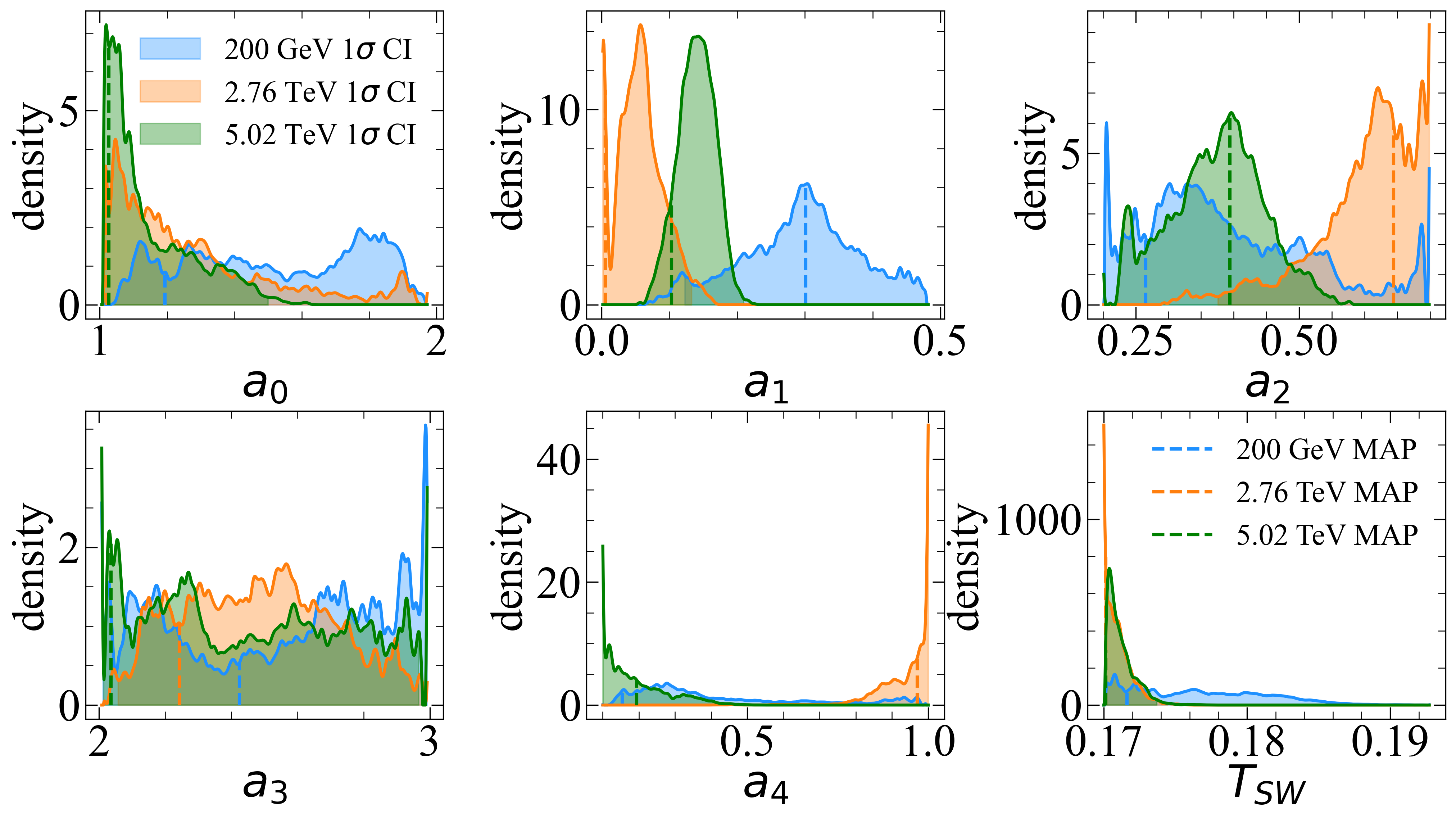}
  \caption{One-dimensional marginals and joint HPD projections for the potential parameters. Different color bands represent three collision systems respectively: \(5.02~\mathrm{TeV}\) Pb–Pb, \(2.76~\mathrm{TeV}\) Pb–Pb, and \(200~\mathrm{GeV}\) Au–Au. The marginal density estimated from SGLD samples are plotted, and the dashed lines mark the Maximum A Posteriori (MAP) Estimation of each parameter for each collision system.  Top \(68.27\%\) of the retained SGLD draws are plotted as an estimation of $1\sigma$ CI for each potential parameter. 
}
  \label{fig:param-seperate}
\end{figure}

To determine the $1\sigma$ uncertainty, we identify the $68.27\%$ of the SGLD samples that yield the minimal loss, representing the Highest Posterior Density (HPD) region, displayed in 
Figure \ref{fig:param-seperate}. Additionally, the optimal parameter values are indicated by the dashed lines in each panel. The three colors distinguish the heavy quark potential configurations corresponding to the three different collision systems.

The $1\sigma$ credible intervals and the Maximum A Posteriori values for each potential parameter are summarized in Table \ref{lab-tab:extract-V-method1}. Notably, the MAP value of $a_4$, which governs the magnitude of the Debye mass, is found to be much smaller than 1.0 for both 5.02 TeV Pb-Pb and 200 GeV Au-Au collisions. This suggests that the color screening effect is only weakly manifested in the real part of the heavy quark potential. In contrast, at 2.76 TeV, the extracted value of $a_4$ is approximately 1.0, representing a significant discrepancy compared to the other two collision systems. The switching temperature $T_{\rm sw}$, defined as the point where the in-medium potential transitions to the vacuum Cornell potential, is extracted to be remarkably close to the critical temperature $T_c$. This finding indicates that the current parametrization of $V(T, r)$ is sufficient to describe bottomonium observables across all three collision systems. Consequently, there is no practical necessity to employ a hybrid form that explicitly combines the parametrized complex potential with the vacuum Cornell potential.

\begin{table}[t]
\centering
\setlength{\tabcolsep}{3pt}      % 仅对本表生效
\renewcommand{\arraystretch}{1.15}
\small
\caption{Optimal values and $1\sigma$ credible intervals of the potential parameters presented separately at three collision energies.}
\label{lab-tab:extract-V-method1}
\begin{tabular}{|c|c|c|c|c|}
\hline
\textbf{Parameter} &  \textbf{5.02 TeV Pb--Pb} & \textbf{2.76 TeV Pb--Pb} & \textbf{200 GeV Au--Au} & \textbf{overlap}\\
\hline
$a_0$  & $[1.005,\,1.500]$ & $[1.011,\,1.941]$ & $[1.039,\,1.971]$ & $[1.039,\,1.500]$\\
$a_1$  & $[0.051,\,0.209]$ & $[0.002,\,0.132]$ & $[0.122,\,0.47]$ & $[0.122,\,0.132]$ \\
$a_2$  & $[0.221,\,0.560]$ & $[0.375,\,0.698]$ & $[0.201,\,0.692]$ & $[0.375,\,0.560]$\\
$a_3$  & $[2.012,\,2.967]$ & $[2.056,\,2.983]$ & $[2.01,\,2.992]$ & $[2.056,\,2.967]$\\
\hline
$a_4$  & $[0.100,\,0.497]$ & $[0.670,\,1]$ & $[0.120,\,0.968]$ & None\\
\hline
$T_{\rm sw}$  & $[0.170,\,0.173]$ & $[0.170,\,0.174]$ & $[0.170,\,0.185]$ & $[0.1701,\,0.173]$\\
\hline
\multicolumn{5}{|l|}{\textbf{Potential parametrization:}} \\
\multicolumn{5}{|l|}{%
  \shortstack[l]{%
  $\displaystyle V_I = -\,i\,T^{a_{0}} \bigl(a_1 r + a_2 r^{a_3}\bigr),$\\
  $\displaystyle V_R = -\alpha\!\left(\mu_D + \frac{e^{-\mu_D r}}{r}\right)
   + \frac{2\sigma}{\mu_D}
   - \frac{\sigma e^{-\mu_D r}\,(2 + \mu_D r)}{\mu_D},\quad
   \mu_D = a_4\,T \sqrt{\frac{4\pi N_c}{3}\!\left( 1 + \frac{N_f}{6} \right)\alpha}\,.$
  }%
} \\
\hline
\multicolumn{5}{|l|}{\textbf{MAP values of $\bf(a_0,a_1,a_2,a_3,a_4,T_{\rm sw})$:}} \\
\multicolumn{5}{|l|}{%
  \shortstack[l]{%
5.02~TeV Pb--Pb: $(1.026,\,0.103,\,0.393,\,2.035,\,0.193,\,0.170)$;\;\\
2.76~TeV Pb--Pb: $(1.021,\,0.005,\,0.644,\,2.242,\,0.968,\,0.170)$;\;\\
200~GeV Au--Au: $(1.192,\,0.300,\,0.265,\,2.42,\,0.153,\,0.171)$.
  }%
} \\
\hline
\end{tabular}
\end{table}

\begin{figure}[htbp]
  \centering
\includegraphics[width=0.8\textwidth]{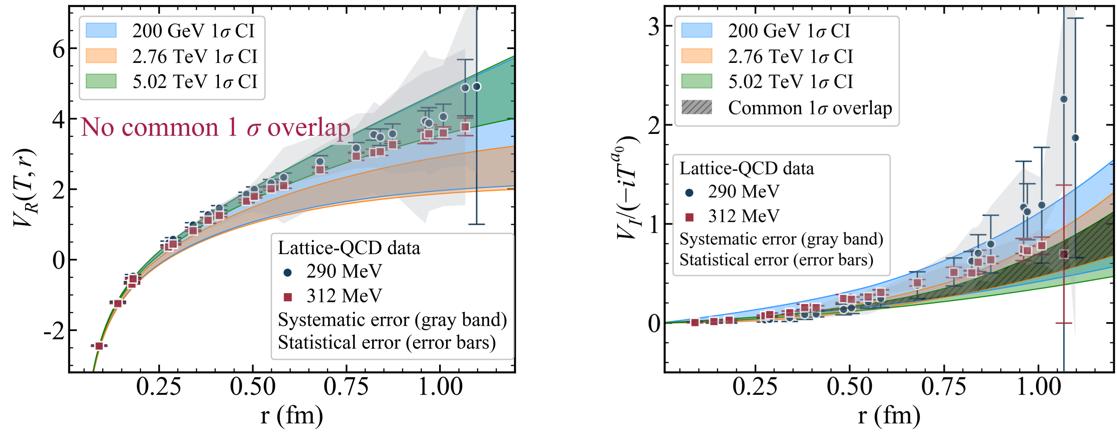}
\caption{The $1\sigma$ credible intervals for the in-medium heavy quark potentials, extracted via the CNN-SGLD framework across three collision systems (5.02 TeV Pb-Pb, 2.76 TeV Pb-Pb, and 200 GeV Au-Au), are depicted by distinct colored bands at $T=300$ MeV.
The real and imaginary parts of the potentials are displayed in the left and right panels, respectively. Lattice QCD results are from Ref.~\cite{Burnier_2017}.
}
  \label{fig:overlap-bands}
\end{figure}

Based on the values provided in Table \ref{lab-tab:extract-V-method1}, the optimal parametrizations for the real and imaginary parts of the heavy quark potentials at the three collision energies are shown in Figure  \ref{fig:overlap-bands}. The $1\sigma$ confidence intervals (CIs) for each potential are indicated by distinct colored bands. An overlap among the three uncertainty bands is clearly observable; specifically, the three imaginary potentials demonstrate high consistency. While the real-part potentials exhibit some discrepancies, these are primarily attributed to variations in the extracted Debye masses across the different collision systems. Consequently, the extracted imaginary parts of the heavy quark potential remain consistent across the three collision systems. In contrast, the real parts exhibit certain discrepancies, highlighting the necessity for a universal extraction of the potentials by simultaneously incorporating bottomonium $R_{AA}$ data from various collision energies, as implemented in the main text.

%%%%%%%%%%%%%%%%%%
%%%%%%%%%%%%%%%%%%

\end{appendices}

%\appendix
\vspace{2cm}
\acknowledgments
JL and BC are supported by the National Natural Science Foundation of China
(NSFC) under Grant Nos. 12575149, 12175165. KZ is supported by the NSFC under Grant Nos. 92570117, the CUHK-Shenzhen University development fund under grant No. UDF01003041 and UDF03003041, and Shenzhen Peacock fund under No. 2023TC0179, Ministry of Science and Technology of China under Grant No. 2024YFA1611004.

%This is the most common positions for acknowledgments. A macro is
%available to maintain the same layout and spelling of the heading.

%\paragraph{Note added.} This is also a good position for notes added
%after the paper has been written.

%\bibliographystyle{unsrt}
\bibliographystyle{JHEP}
\bibliography{ref}

%\begin{thebibliography}{99}

%\bibitem{Choi:20} review on SSE:  J. Phys.: Condens. Matter 24 (2012) 273201 

% Please avoid comments such as "For a review'', "For some examples",
% "and references therein" or move them in the text. In general,
% please leave only references in the bibliography and move all
% accessory text in footnotes.

% Also, please have only one work for each \bibitem.
%\end{thebibliography}

\end{document}